\documentclass[11pt]{article}

\usepackage[preprint]{acl}

\usepackage{times}
\usepackage{latexsym}

\usepackage[T1]{fontenc}

\usepackage[utf8]{inputenc}

\usepackage{microtype}

\usepackage{inconsolata}

\usepackage{graphicx}

\usepackage{enumitem}
\usepackage{pifont}
\usepackage{makecell}
\usepackage[framemethod=TikZ]{mdframed}
\usepackage{multirow}
\usepackage{anyfontsize}
\usepackage{adjustbox}
\usepackage[labelfont=bf,textfont=md]{caption}
\usepackage{rotating}
\usepackage{xcolor}
\usepackage{hhline}
\usepackage{threeparttable}
\usepackage{xspace}
\usepackage{appendix}
\usepackage{colortbl}
\usepackage{booktabs}
\usepackage{tablefootnote}

\newcommand*\rot{\rotatebox{90}}

\newcommand{\summary}[1]{\textsc{Experiment {#1} Summary:}}

\definecolor{myblue}{RGB}{29, 105, 150}
\definecolor{myorange}{RGB}{225, 124, 5}

\setlist{noitemsep,topsep=0pt,parsep=0pt,partopsep=0pt}

\newtheorem{theorem}{Prompt}

\newcommand{\cmark}{\ding{51}}
\newcommand{\xmark}{\ding{55}}

\newcommand{\python}{\texttt{Python}\xspace}

\definecolor{mygray}{RGB}{247,247,247}

\usepackage{tikzpagenodes}
\usepackage{tikz}

\title{A Study of LLMs' Preferences for Libraries and Programming Languages}

\author{
  \textbf{Lukas Twist\textsuperscript{1}} \quad
  \textbf{Mark Harman\textsuperscript{2}} \quad
  \textbf{Don Syme\textsuperscript{3}} \\
  \textbf{Joost Noppen\textsuperscript{1,4}} \quad
  \textbf{Helen Yannakoudakis\textsuperscript{1}} \\
  \textbf{Detlef Nauck\textsuperscript{4}} \quad
  \textbf{Jie M. Zhang\textsuperscript{1}}
\\
\\
  \textsuperscript{1}King's College London, London, UK \\
  \textsuperscript{2}University College London, London, UK \\
  \textsuperscript{3}GitHub Next, London, UK \\
  \textsuperscript{4}Digital AI Research, BT Group, Ipswich, UK
}

\begin{document}

\maketitle

\begin{tikzpicture}[remember picture, overlay]
\node[anchor=south, yshift=15pt] at (current page.south) {
\parbox{0.9\textwidth}{
\centering
\footnotesize
\textit{Accepted to Findings of the Association for Computational Linguistics: ACL 2026.}
}
};
\end{tikzpicture}

\begin{abstract}

Despite the rapid progress of large language models (LLMs) in code generation, existing evaluations focus on functional correctness or syntactic validity, overlooking how LLMs make critical design choices such as which library or programming language to use.
To fill this gap, we perform the first systematic study of LLMs' preferences for libraries and programming languages when generating code, covering eight diverse LLMs.
We observe a strong tendency to overuse widely adopted libraries such as \texttt{NumPy}; in up to 45\% of cases, this usage is not required and deviates from the ground-truth solutions.
The LLMs we study also show a significant preference toward \texttt{Python} as their default language.
For high-performance project initialisation tasks where using \texttt{Python} may bring more security and efficiency risks, it remains the dominant choice in 58\% of cases, and \texttt{Rust} is not used once.
These results highlight how LLMs prioritise familiarity and popularity over suitability and task-specific optimality;
underscoring the need for targeted fine-tuning, data diversification, and evaluation benchmarks that explicitly measure language and library selection fidelity.

\end{abstract}

\section{Introduction}

Large Language Models (LLMs) have recently made rapid progress~\cite{naveedComprehensiveOverviewLarge2024}, particularly excelling in code generation~\cite{jiangSurveyLargeLanguage2024}, with extensive work on evaluating and improving LLM code accuracy, security, and efficiency~\cite{chenSurveyEvaluatingLarge2024}.
However, less studied are the \textit{preferences} LLMs exhibit when choosing the libraries and programming languages to use for a coding task.
These preferences can have tangible effects on software performance, efficiency, maintainability, and ecosystem diversity --- potentially reinforcing dominant technologies and overlooking domain-suitable alternatives.
Understanding these tendencies is therefore crucial for both building trustworthy and context-aware code generation systems, and for designing benchmarks that capture not only correctness but also the rationale behind design choices in LLM-generated code.



To address this gap, we present the first systematic study of library and programming language preferences across eight production-grade LLMs.
We begin with libraries -- vital components of modern software development~\cite{somervilleSoftwareEngineeringGlobal2016} -- because developers often omit them from prompts~\cite{haoEmpiricalStudyDevelopers2024} and may not know which dependencies a task requires~\cite{larios-vargasSelectingThirdpartyLibraries2020}, giving LLMs the opportunity to influence library adoption.
We then consider the programming language preferences of LLMs.
Many end users lack the expertise to judge whether an LLM’s choice of language is appropriate~\cite{khuranaWhyWhenLLMBased2024,nguyenHowBeginningProgrammers2024} and leave the decision entirely to the LLM, especially in end-to-end automatic software generation scenarios~\cite{sarkarVibeCodingProgramming2025}.

For both libraries and programming languages, we examine LLM preferences in two key scenarios: writing code for practical tasks from widely studied benchmarks; and generating the initial code for new projects, a crucial phase where LLMs have been found to be particularly helpful~\cite{rasnayakaEmpiricalStudyUsage2024} and foundational technology choices are made.
In addition, we explore whether the technologies that LLMs recommend in natural language (NL) responses match those they actually use when generating code.

Our experiments lead to the following primary observations:
\textbf{1)}~All LLMs we study \textbf{heavily favour well established libraries over high-momentum alternatives}.
In particular, LLMs favour over-using widely adopted data science libraries even when not required for the task; for example, \texttt{NumPy} usage diverges with ground-truth solutions for up to 45\% of tasks.
\textbf{2)}~For programming languages, we observe a \textbf{significant preference towards using \texttt{Python}}.
To our surprise, even for tasks where high performance and memory safety are critical -- and therefore \texttt{Python} is considered suboptimal -- it remains the most used language in 58\% of cases.
\textbf{3)}~\textbf{LLMs do not follow their own recommendations}: the technologies suggested in NL responses are a weak signal of those actually used in generated code.
These tendencies likely arise from multiple factors, such as the prevalence of certain libraries in open source repositories, the distribution of programming languages included in training data~\cite{wangExploringMultiLingualBias2024}, or post-training alignment and fine-tuning~\cite{zieglerFineTuningLanguageModels2020}.

Such preferences have mixed consequences.
On the positive side, defaulting to well-known libraries and languages can speed prototyping and increase interoperability for many users.
On the negative side, persistent favouritism can \textbf{hurt the open-source community} by marginalizing emerging projects, ultimately triggering \textbf{a vicious cycle} with less and less diversity as LLMs evolve with self-generated data~\cite{zhuHowSynthesizeText2025}. Most critically, these biases can present a significant \textbf{security threat}, potentially steering users toward suboptimal and insecure implementation choices.
Understanding the extent of these preferences is important, not only for assessing the potential risks and broader impacts of LLMs on the software ecosystem, but also for guiding LLM creators in improving model design, training data composition, and alignment strategies to achieve more balanced and context-aware code generation.


\noindent\textit{Our contributions are as follows.}
\begin{enumerate}[left=0pt]
    \item We present the first systematic study of LLM preferences for libraries and programming languages when generating code.
    \item We highlight the importance of analysing and quantifying diversity and preference patterns in LLM-generated code, and discuss their potential implications for software development practices and the open-source ecosystem.
    \item We release our code, datasets and complete results publicly via our GitHub repository, to encourage further investigation and improvement in this area:
    \url{https://github.com/itsluketwist/llm-code-bias}
\end{enumerate}


\section{Related Work}

\paragraph{Bias and unfairness in LLMs.}
Bias in the NL outputs of LLMs is well documented: LLMs often inherit and amplify social and representational stereotypes from their training data~\cite{benderDangersStochasticParrots2021a,nadeemStereoSetMeasuringStereotypical2021,gallegosBiasFairnessLarge2024}, and can preferentially surface newer content in answers~\cite{fangLargeLanguageModels2025a}.
Social biases also appear in LLM generated code, and several datasets and frameworks now quantify code-based unfairness across demographics~\cite{liuUncoveringQuantifyingSocial2023,huangdongBiasTestingMitigation2024,lingBiasUnveiledInvestigating2025}.
In contrast, \textit{technological} biases -- preferences for the specific technologies used in the code -- are less explored.
Recent work exposing provider bias (e.g., favouring specific cloud services) motivates our focus on whether analogous biases exist at the library and programming language level~\cite{zhangInvisibleHandUnveiling2025}.

\paragraph{LLM code generation.}
Code generation in LLMs has been extensively studied~\cite{jiangSurveyLargeLanguage2024}, including their ability to implement code using external or unseen libraries~\cite{liuCodeGen4LibsTwoStageApproach2023,zanPrivateLibraryOrientedCodeGeneration2023,patelEvaluatingInContextLearning2024}.
However, evaluation methodology remains an open problem~\cite{paulBenchmarksMetricsEvaluations2024a}.
Existing approaches typically use datasets of coding tasks, but they are often limited in scope and favour popular programming languages such as \python and \texttt{Java}~\cite{yadavPythonSagaRedefiningBenchmark2024a}.
These benchmarks are also often compromised via data leakage -- where the benchmarks' tasks are included in the LLM training data -- giving unfair representations of the LLMs' abilities to complete the tasks~\cite{mattonLeakageCodeGeneration2024, zhouLessLeakBenchFirstInvestigation2025}.
LLM code-generation also raises several ethical issues around code reuse and provenance~\cite{germanCodeSiblingsTechnical2009}, as they are often trained on large public code corpora~\cite{chenEvaluatingLargeLanguage2021}.
Together, these issues motivate our investigation into whether LLM coding preferences, and whether they prefer certain technologies due to their exposure to code during training.

\paragraph{LLM recommendations \& library selection.}
Library selection is a core developer skill that requires trade-offs between functionality, performance and maintainability ~\cite{larios-vargasSelectingThirdpartyLibraries2020,tanzilHowPeopleDecide2024a}.
Increasingly, users turn to LLMs for programming recommendations, with observable declines in public Q\&A activity following the release of ChatGPT~\citep{delrio-chanonaLargeLanguageModels2024,zhongCanLLMReplace2024}.
Furthermore, there is a growing body of work that treats library recommendation as a core capability of LLMs.
ToolLLM~\cite{qinToolLLMFacilitatingLarge2023} and Gorilla~\cite{NEURIPS2024_e4c61f57} show how retrieval-aware or tool-augmented LLMs can recommend and produce API calls reliably;
while APIGen~\cite{chenAPIGenGenerativeAPI2024b} studies generative approaches for recommending APIs.
These prior efforts focus on improving recommendation accuracy; they do not consider how potential biases in LLM recommendations affect developer choices.
This gap motivates our study of whether LLM coding recommendations align with the technologies they actually use.


\section{Experimental Design}

\subsection{LLM Selection}
\label{sec:llm}

We use a diverse set of LLMs in this study, to gain a broad understanding of LLM preferences.
LLMs are chosen to vary in number of parameters, availability (open- or closed-source), and intended use case (general or code-specific).
Therefore, we chose the following eight LLMs for our study: GPT-4o-mini and GPT-3.5-turbo~\cite{openaiModelsOpenAIAPI2025}, Claude-3.5 Sonnet and Haiku~\cite{anthropicClaude3Model2024}, Llama-3.2-3B~\cite{Llama32Model2025}, Mistral-7B~\cite{jiangMistral7B2023}, Qwen-2.5-Coder~\cite{huiQwen25CoderTechnicalReport2024} and DeepSeek-LLM~\cite{deepseek-aiDeepSeekLLMScaling2024}.

To reflect the typical usage of LLMs by developers, which often overlooks the role of parameters~\cite{donatoStudyingHowConfigurations2025}, each LLM is prompted using the default parameter values from its corresponding API.
Furthermore, we conduct each LLM interaction in a fresh API session to avoid bias from prompt caching or leakage~\cite{gu2025auditing};
and we do not use a system prompt to ensure that each LLM has its base functionality considered without external influence~\cite{muCloserLookSystem2025}.

\paragraph{}
\textit{Full details on LLMs, and how we extract data from their responses, are given in Appendix~\ref{app:llm}.}

\subsection{Experiment 1: Library Preferences}

We focus on a single programming language to enable for a more in-depth analysis.
\python is chosen due to its vast collection of open-source libraries~\cite{pypiPyPIPythonPackage2025}, its straightforward import syntax, and its popularity in the open-source community~\cite{staffOctoverseAILeads2024a}.
Previous work shows that users often start tasks with NL descriptions without knowing which -- or that any -- libraries are required~\cite{kuhnDesigningBetterTools2012}; and when prompting LLMs, will initially omit details, iteratively refining their requirements over time~\cite{haoEmpiricalStudyDevelopers2024}.
Therefore, our goal is to measure which external libraries LLMs prefer when prompts do not specify library names.

\paragraph{Benchmark Tasks.}
\label{sub:exp-lib-bench}

We use BigCodeBench \cite{zhuoBigCodeBenchBenchmarkingCode2024} as our primary benchmark.
BigCodeBench contains 1,140 \python tasks across seven domains (general, computation, visualisation, system, time, network, cryptography) --- 813 of which include external libraries in their ground-truth solution.
The latest version is released in February 2025, postdating our LLMs’ knowledge cut-offs, reducing exposure bias.
We use the NL description from each task, and to eliminate bias we filter out any tasks that contain a reference to an external library used in its ground-truth solution.
The resulting dataset contains 525 tasks, using 34 distinct libraries in their ground-truth solutions.

We use a BigCodeBench inspired prompt, asking the model to produce self-contained \python code along with the directive to use an external library (forcing the use of an external library is a deliberate design choice to focus this experiment on library selection).
We generate three independent responses per task to reduce the inherent randomness in LLM outputs~\cite{sallouBreakingSilenceThreats2024}.

\paragraph{Project initialisation tasks.}
\label{sub:exp-lib-proj}

LLMs have been shown to be particularly helpful in the early stages of software projects~\cite{rasnayakaEmpiricalStudyUsage2024};
this is also a natural decision point for adopting a new library, where the LLM's preferences may influence users' choices.
Therefore, to complement benchmark tasks, we also measure library preferences in a more open-ended, project-level setting.

We create five realistic project descriptions inspired by categories on Awesome Python~\cite{pythonAwesomePython2025}, a curated collection of \python libraries, to ensure that multiple viable libraries exist for each task.
We intentionally use open-ended prompts (allowing the LLM to use one or more libraries), requesting the LLMs ``write the initial code'' for the following types of projects: \textit{database}, \textit{deep learning}, \textit{distributed computing}, \textit{web scraper}, and \textit{web server}.
For each project description, we generate 100 responses per LLM and analyse the distribution of libraries used to mitigate the variance of the evaluation~\cite{madaanQuantifyingVarianceEvaluation2024}.

\paragraph{}
\textit{Complete dataset details and full prompts for Experiment 1 are given in Appendix~\ref{app:exp1}.}

\subsection{Experiment 2: Language Preferences}
\label{sub:exp-lang}

Next, we investigate LLM preferences for programming languages when the prompts do not specify a target language.
Although this is less common among professional developers -- mostly seen from novice programmers who struggle to write detailed prompts~\cite{nguyenHowBeginningProgrammers2024}, and “vibe-coders”~\cite{smithAIVibeCoding2025}, that leave more of the technical decisions to the LLM -- it will allow us to further study the inherent biases LLMs exhibit when allowed to make coding decisions.

\paragraph{Benchmark Tasks.}
\label{sub:exp-lang-bench}

We use language-agnostic datasets, chosen to (i) contain NL-only descriptions and (ii) reduce single-language bias or leakage by having published solutions in multiple languages.
We use six widely-adopted datasets, spanning three categories: \textit{basic} (Multi-HumanEval, MBXP), \textit{real-world} (AixBench, CoNaLa), and \textit{coding challenge} (APPS, CodeContests).
Any task whose description explicitly mentions a programming language from the TIOBE Index~\cite{jansenTIOBEIndex2025} (top 50) is removed.
Preliminary experiments showed that a sample of 200 tasks gave a good representation of the results for the larger datasets, therefore if a dataset was over 200 tasks, it would be sampled, allowing results to be fairly aggregated.

We use a lightweight prompt template compatible with each dataset, asking for a code solution and explanation; we request an explanation because we found this to be the most reliable way to encourage the LLM to respond using \texttt{Markdown} notation without biasing the results.
We generate three independent responses per task to reduce the inherent randomness in LLM outputs~\cite{sallouBreakingSilenceThreats2024}.

\paragraph{Project Initialisation Tasks.}
\label{sub:exp-lang-proj}

We also want to investigate LLM preferences for programming languages when asked to write initial project code.
Our initial results imply that LLMs have a general preference towards using \python.
Therefore, to challenge this default, we choose project descriptions with non-functional requirements such that \python is regarded as a suboptimal choice.
Although \python offers user-friendly syntax and rapid prototyping, it is less suitable for high-performance workloads due to its interpretive nature and inefficient concurrency implementation~\cite{gavrilovaProsConsPython2023}.
For computationally intensive tasks, compiled languages like \texttt{C}, \texttt{C++}, and \texttt{Rust} are generally preferred; they provide greater memory control, lower execution overhead, and true parallel processing~\cite{costanzoPerformanceVsProgramming2021}.

We prompt the LLM to ``write the initial code'' for the following project descriptions, each representative of a domain where \python is suboptimal: \textit{concurrent web server}, \textit{cross-platform graphical user interface}, \textit{low-latency trading platform}, \textit{parallel task processing library}, and a \textit{system-level application}.
The descriptions are intentionally open-ended, allowing the use of multiple languages if required.
For each project description, we generate 100 responses per LLM and analyse the distribution of programming languages used to mitigate the variance of the evaluation~\cite{madaanQuantifyingVarianceEvaluation2024}.

\paragraph{}
\textit{Complete dataset details and full prompts for Experiment 2 are given in Appendix~\ref{app:exp2}.}

\subsection{Experiment 3: Recommendation Consistency}
\label{sub:exp-consistency}

Finally, we want to examine whether the LLMs' NL recommendations for libraries and programming languages match what they actually use when generating code.
This directly tests whether the LLM is self-consistent with its internal knowledge of what technology to use for what task, or if it has bias specific to code generation.
To obtain the NL recommendations, we prompt the LLMs to ``list, in order'' the best \python library or programming language to use for each project initialisation task from Experiments 1 and 2.

We generate three responses per task and take the arithmetic mean of the ranks to obtain the final ranking per LLM.
We then derive empirical usage rankings from the results of Experiments 1 and 2.
For each task, we rank the libraries or programming languages descending by their observed usage frequencies.
We measure correspondence between the two rankings with Kendall's $\tau_b$ coefficient, a non-parametric (no assumptions about the underlying distribution of the data) rank correlation suitable for ordinal data (natural order but not necessarily equal intervals) that is robust to ties~\cite{kendallRankCorrelationMethods1990}.

\paragraph{}
\textit{Complete prompt selection and calculation details for Experiment 3 are given in Appendix~\ref{app:exp3}.}


\begin{figure*}
    \centering
    \includegraphics[width=1.0\textwidth]{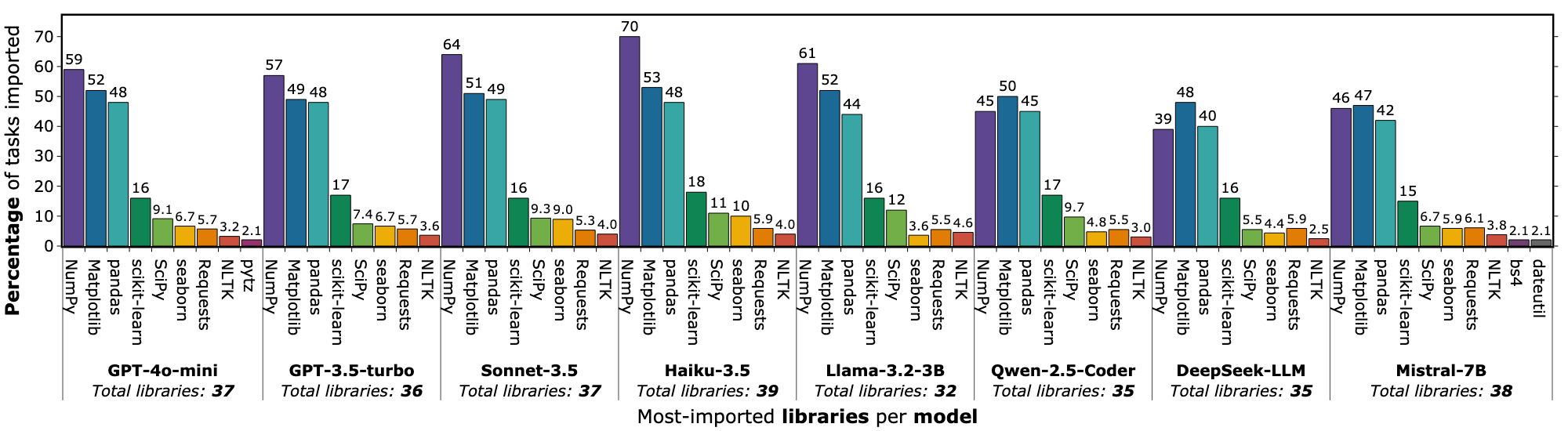}
    \caption{\textbf{\textit{Experiment 1: Library Preferences for Benchmark Tasks.}} Libraries used by LLMs when responding to BigCodeBench tasks. For each LLM, we give the most-used libraries with the percentage of tasks that had a response importing them, and the total unique libraries used. Other libraries are imported for \textit{less than 2\%} of tasks.}
    \label{fig:library-problems}
\end{figure*}





\section{Results}\label{sec:results}

Here we present the results.
\textit{Note} that it is expected that the percentages in the results do not add up to 100\% because LLMs were prompted multiple times per task and could respond with different libraries or languages, and sometimes without code.

\subsection{Experiment 1: Library Preferences}\label{sub:rq2}

This experiment aims to assess LLMs' preferences for \texttt{Python} libraries during code generation.

\paragraph{Benchmark tasks.} \label{subsub:lib-div-prob}

Figure~\ref{fig:library-problems} shows the libraries used by LLMs when writing code for BigCodeBench tasks.
All LLMs produce very similar distributions: the top three libraries are identical (\texttt{NumPy}, \texttt{pandas}, \texttt{Matplotlib}), with a clear gap between them and the next most-used library.
Although a portion of BigCodeBench focuses on computation and visualisation -- domains where these libraries are naturally useful -- we observe that the LLMs frequently use them when the ground truth does not.
For example, \texttt{NumPy} is used in responses to 192 of the 305 tasks where it is not part of the ground-truth solution, indicating a tendency to use it even when not required for the given task.
\textit{Further ground-truth analysis is provided in Appendix~\ref{app:res1}.}

\texttt{Python}’s ecosystem is huge: in July 2025 over 7,000 libraries surpassed 100,000 downloads~\cite{TopPyPIPackages2025} (still only a fraction of those available).
However, each LLM in our study used surprisingly few distinct libraries (32-39) across hundreds of varied tasks.
Although not all of the available libraries are suitable, using less than 40 points to a lack of diversity in the libraries that LLMs choose when not given a specific directive.

\paragraph{Prompt sensitivity}
We use a prompt that explicitly requests an external library to focus our study on library selection.
We perform a small ablation study to assess LLM sensitivity to this choice, repeating the experiment with prompt variations that express different levels of strictness around external library usage.
The results show that our main findings are robust, with similar library preferences observed across all prompt variants.
\textit{Alternative prompts and full results for the ablation study can be seen in Appendix~\ref{app:res-ablation}.}

\begin{table*}[ht]
    \caption{\textbf{\textit{Experiment 1: Library Preferences for Project Initialisation Tasks.}} Libraries used by LLMs when writing initial project code. The libraries (\textit{l}) used by each LLM are given, along with the percentage (\textit{p}) of responses that used that library. Libraries considered to provide core functionality are marked with a \textbf{*} and are listed first.}
    \label{tab:library-projects}
    
    \centering
    \begin{threeparttable}
    \begin{adjustbox}{width=\textwidth}
    \bgroup
    \def\arraystretch{1.1}

\begin{tabular}{clrlrlrlrlrlrlrlr}
\toprule
\multirowcell{2}{\textbf{Library}\\\textbf{Task}} & \multicolumn{2}{c}{\textbf{GPT-4o-mini}} & \multicolumn{2}{c}{\textbf{GPT-3.5-turbo}} & \multicolumn{2}{c}{\textbf{Sonnet-3.5}} & \multicolumn{2}{c}{\textbf{Haiku-3.5}} & \multicolumn{2}{c}{\textbf{Llama-3.2-3B}} & \multicolumn{2}{c}{\textbf{Qwen-2.5-Coder}} & \multicolumn{2}{c}{\textbf{DeepSeek-LLM}} & \multicolumn{2}{c}{\textbf{Mistral-7B}} \\
\cmidrule(lr){2-3}\cmidrule(lr){4-5}\cmidrule(lr){6-7}\cmidrule(lr){8-9}\cmidrule(lr){10-11}\cmidrule(lr){12-13}\cmidrule(lr){14-15}\cmidrule(lr){16-17}
 & \makecell{\textit{l}} & \multicolumn{1}{c}{\textit{p}}& \makecell{\textit{l}} & \multicolumn{1}{c}{\textit{p}}& \makecell{\textit{l}} & \multicolumn{1}{c}{\textit{p}}& \makecell{\textit{l}} & \multicolumn{1}{c}{\textit{p}}& \makecell{\textit{l}} & \multicolumn{1}{c}{\textit{p}}& \makecell{\textit{l}} & \multicolumn{1}{c}{\textit{p}}& \makecell{\textit{l}} & \multicolumn{1}{c}{\textit{p}}& \makecell{\textit{l}} & \multicolumn{1}{c}{\textit{p}} \\ \midrule
\makecell{Database} & \makecell[l]{SQLAlchemy*\\models\\database\\-} & \makecell[r]{100\%\\11\%\\3\%\\-} & \makecell[l]{SQLAlchemy*\\-\\-\\-} & \makecell[r]{100\%\\-\\-\\-} & \makecell[l]{SQLAlchemy*\\-\\-\\-} & \makecell[r]{100\%\\-\\-\\-} & \makecell[l]{SQLAlchemy*\\Pydantic\\dotenv\\\textit{Total used}} & \makecell[r]{100\%\\3\%\\3\%\\\textit{5}} & \makecell[l]{SQLAlchemy*\\models\\db\\-} & \makecell[r]{100\%\\96\%\\1\%\\-} & \makecell[l]{SQLAlchemy*\\TensorFlow\\-\\-} & \makecell[r]{99\%\\1\%\\-\\-} & \makecell[l]{SQLAlchemy*\\-\\-\\-} & \makecell[r]{100\%\\-\\-\\-} & \makecell[l]{SQLAlchemy*\\-\\-\\-} & \makecell[r]{100\%\\-\\-\\-} \\ \midrule
\makecell{Deep learning} & \makecell[l]{TensorFlow*\\scikit-learn*\\PyTorch*\\TorchVision*\\NumPy\\Matplotlib} & \makecell[r]{97\%\\4\%\\3\%\\2\%\\76\%\\64\%} & \makecell[l]{TensorFlow*\\PyTorch*\\Keras*\\scikit-learn*\\TorchVision*\\NumPy} & \makecell[r]{85\%\\7\%\\6\%\\2\%\\1\%\\47\%} & \makecell[l]{PyTorch*\\TorchVision*\\scikit-learn*\\Matplotlib\\NumPy\\-} & \makecell[r]{100\%\\93\%\\5\%\\96\%\\11\%\\-} & \makecell[l]{scikit-learn*\\TensorFlow*\\PyTorch*\\TorchVision*\\NumPy\\Matplotlib} & \makecell[r]{82\%\\75\%\\25\%\\14\%\\86\%\\26\%} & \makecell[l]{TensorFlow*\\scikit-learn*\\NumPy\\Matplotlib\\-\\-} & \makecell[r]{100\%\\91\%\\100\%\\91\%\\-\\-} & \makecell[l]{TensorFlow*\\-\\-\\-\\-\\-} & \makecell[r]{100\%\\-\\-\\-\\-\\-} & \makecell[l]{TensorFlow*\\-\\-\\-\\-\\-} & \makecell[r]{100\%\\-\\-\\-\\-\\-} & \makecell[l]{Keras*\\scikit-learn*\\NumPy\\-\\-\\-} & \makecell[r]{100\%\\44\%\\100\%\\-\\-\\-} \\ \midrule
\makecell{Distributed\\computing} & \makecell[l]{Dask*\\MPI4py*\\Ray*\\NumPy\\\textit{Total used}} & \makecell[r]{73\%\\5\%\\2\%\\28\%\\\textit{8}} & \makecell[l]{MPI4py*\\Dask*\\NumPy\\Joblib\\-} & \makecell[r]{33\%\\31\%\\6\%\\1\%\\-} & \makecell[l]{Dask*\\Ray*\\Celery*\\Redis\\\textit{Total used}} & \makecell[r]{24\%\\20\%\\1\%\\49\%\\\textit{8}} & \makecell[l]{Dask*\\Ray*\\Celery*\\NumPy\\\textit{Total used}} & \makecell[r]{89\%\\4\%\\2\%\\71\%\\\textit{6}} & \makecell[l]{Dask*\\NumPy\\-\\-\\-} & \makecell[r]{100\%\\55\%\\-\\-\\-} & \makecell[l]{Ray*\\-\\-\\-\\-} & \makecell[r]{100\%\\-\\-\\-\\-} & \makecell[l]{-\\-\\-\\-\\-} & \makecell[r]{-\\-\\-\\-\\-} & \makecell[l]{Dask*\\NumPy\\-\\-\\-} & \makecell[r]{100\%\\100\%\\-\\-\\-} \\ \midrule
\makecell{Web scraper} & \makecell[l]{BS4*\\Requests*\\pandas\\-} & \makecell[r]{100\%\\100\%\\27\%\\-} & \makecell[l]{BS4*\\Requests*\\pandas\\-} & \makecell[r]{100\%\\100\%\\13\%\\-} & \makecell[l]{BS4*\\Requests*\\pandas\\-} & \makecell[r]{100\%\\100\%\\100\%\\-} & \makecell[l]{BS4*\\Requests*\\pandas\\\textit{Total used}} & \makecell[r]{100\%\\100\%\\100\%\\\textit{6}} & \makecell[l]{BS4*\\Requests*\\pandas\\-} & \makecell[r]{100\%\\100\%\\100\%\\-} & \makecell[l]{BS4*\\Requests*\\pandas\\-} & \makecell[r]{100\%\\100\%\\100\%\\-} & \makecell[l]{BS4*\\Requests*\\-\\-} & \makecell[r]{100\%\\100\%\\-\\-} & \makecell[l]{BS4*\\Requests*\\pandas\\-} & \makecell[r]{100\%\\100\%\\100\%\\-} \\ \midrule
\makecell{Web server} & \makecell[l]{Flask*\\FastAPI*\\Flask-REST\\Flask-Cors\\\textit{Total used}} & \makecell[r]{98\%\\2\%\\19\%\\15\%\\\textit{7}} & \makecell[l]{Flask*\\-\\-\\-\\-} & \makecell[r]{100\%\\-\\-\\-\\-} & \makecell[l]{FastAPI*\\Flask*\\Pydantic\\Uvicorn\\Flask-Cors} & \makecell[r]{48\%\\39\%\\48\%\\48\%\\14\%} & \makecell[l]{Flask*\\FastAPI*\\Flask-Cors\\Flask-SQLA\\\textit{Total used}} & \makecell[r]{69\%\\23\%\\52\%\\31\%\\\textit{12}} & \makecell[l]{Flask*\\Flask-SQLA\\-\\-\\-} & \makecell[r]{100\%\\100\%\\-\\-\\-} & \makecell[l]{Flask*\\-\\-\\-\\-} & \makecell[r]{100\%\\-\\-\\-\\-} & \makecell[l]{Flask*\\-\\-\\-\\-} & \makecell[r]{100\%\\-\\-\\-\\-} & \makecell[l]{Flask*\\-\\-\\-\\-} & \makecell[r]{100\%\\-\\-\\-\\-} \\

\bottomrule
\end{tabular}
    \egroup

    \end{adjustbox}
    
        \vspace{-2mm}
        \begin{tablenotes}
            \item \tiny{\textit{Note some library names have been shortened in the table: Flask-REST is Flask-RESTful, Flask-SQLA is Flask-SQLAlchemy, BS4 is BeautifulSoup}}. 
        \end{tablenotes}
    \end{threeparttable}
\end{table*}

\paragraph{Project initialisation tasks.}

Table~\ref{tab:library-projects} shows the libraries imported when LLMs generate the initial code for five project descriptions.
Open-source LLMs show minimal diversity, repeatedly using the same core libraries across runs, whereas closed-source LLMs use a wider variety of libraries for core functionality.
We observe a correlation between the library diversity and the LLMs' default temperature settings (our open-source LLMs use a default of 0.6--0.7; our closed-source default to 1.0), suggesting that open-source LLMs give developers a more consistent experience.
\textit{We explore the effects of temperature further in Appendix~\ref{app:temperature}.}

We also observe repeated instances of apparent overuse.
For example, Qwen-2.5-Coder imports \texttt{TensorFlow} for the \textit{database} project, and multiple LLMs import \texttt{NumPy} or \texttt{pandas} for \textit{distributed computing}.
These patterns mirror the results for the benchmark tasks, indicating similar preferences in both constrained and open-ended tasks.
Such tendencies may be related to the recent growth in \python that is attributed to its increasing role in AI and data science-based code~\cite{staffOctoverseAILeads2024a}.

\paragraph{Cases analysis.}

\begin{figure}[ht]
    \centering
    \includegraphics[width=\columnwidth]{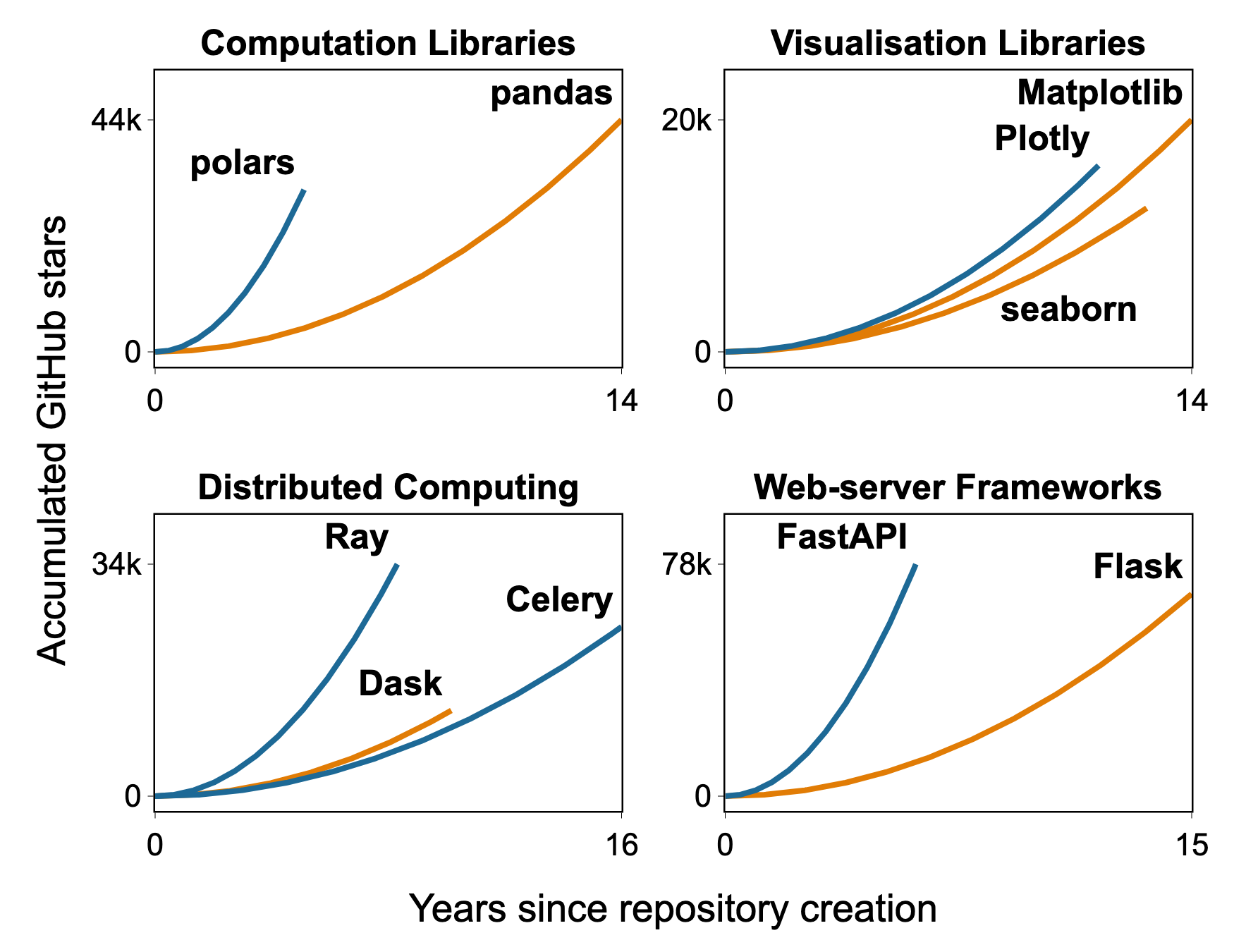}
    \caption{
        \textbf{\textit{Case Analysis.}}
        GitHub growth statistics for libraries studied in case analysis.
        For each library, the accumulated GitHub stars are plotted over its lifetime (data from GitHub, December 2024).
        The top/bottom two graphs show case analysis for benchmark/project initialisation tasks.
        \textcolor{myorange}{\textbf{Orange}} curves indicate libraries that are heavily used in this study, \textcolor{myblue}{\textbf{Blue}} curves indicate libraries with low usage rates.
    }
    \label{fig:case-analysis}
\end{figure}

Our results indicate that LLMs show a preference for established libraries.
To understand the extent of this, we perform a case analysis on libraries used in different domains.
We use GitHub star growth over time as a lightweight proxy for community momentum and look for alternative libraries that show signs of higher potential but have lower usage rates; Figure~\ref{fig:case-analysis} shows this for competing libraries in our sample.
We acknowledge that GitHub stars are an imperfect measure of ecosystem momentum, and complementary indicators -- such as downloads, dependents, or contribution velocity -- could provide a more nuanced view of under-used libraries.
We use star growth because it is widely available, interpretable, and consistently reported across the libraries under study, allowing for uniform comparison.
We find high-momentum alternatives in four domains, showing a pattern of LLMs preferring well-established libraries and underusing newer alternatives.

\begin{enumerate}[left=0pt]

    \item \textit{Computation tasks:} \texttt{pandas} (2010) appears in the 58\% of tasks. \texttt{Polars} (2020) is a newer library with similar functionality and twice as fast GitHub star growth, but is not used at all.
    
    \item \textit{Visualisation tasks:} \texttt{Matplotlib} (2011) and \texttt{seaborn} (2012) are heavily used, for 57\% and 17\% of tasks. \texttt{Plotly} (2013) has additional functionality and higher momentum, but is used for only \textit{one} problem and by \textit{three} LLMs.

    \item \textit{Web server project:} \texttt{Flask} (2010) is used in \textit{88\%} of responses.
    \texttt{FastAPI} (2018) has shown substantially faster recent adoption, yet was used only in 9\% of responses, by only \textit{three} LLMs.

    \item \textit{Distributed computing project:} \texttt{Dask} (2015) is the most common choice, used in 52\% of the responses.
    \texttt{Ray} (2016) and \texttt{Celery} (2009) both have similar functionality along with more stars and faster growth, but they were used minimally (in 16\% and 0.4\% of responses, respectively).
    
\end{enumerate}

\paragraph{\summary{1}}
LLMs show a strong tendency to use older, well-established libraries, often overlooking newer alternatives with high community momentum; as shown in our four case studies, where usage rate differences range from \textbf{36\%} to \textbf{79\%}.
We also observe the overuse of data science libraries: for up to \textbf{45\%} of the benchmark tasks where \texttt{NumPy} is imported, its usage differs from the ground-truth solutions.

\begin{table*}[ht]
    \caption{\textbf{\textit{Experiment 2: Language Preferences for Benchmark Tasks.}} Programming languages used by LLMs when writing code for benchmark tasks. 
    \textit{Dataset} shows the benchmark name, as well as the programming languages for which the ground truth solutions are provided. 
    For each LLM and dataset, the preferred languages (\textit{l}) are given, along with the percentage (\textit{p}) of dataset problems that have a response using that language, and the total count of different languages used (when necessary). The most-used language in each case is in bold.}
    \label{tab:language-problems}
    
    \centering
    \begin{threeparttable}
    \begin{adjustbox}{width=\textwidth}
    \bgroup
    \def\arraystretch{1.1}

\begin{tabular}{clrlrlrlrlrlrlrlr}
\toprule
\multirowcell{2}{\textbf{Dataset}} & \multicolumn{2}{c}{\textbf{GPT-4o-mini}} & \multicolumn{2}{c}{\textbf{GPT-3.5-turbo}} & \multicolumn{2}{c}{\textbf{Sonnet-3.5}} & \multicolumn{2}{c}{\textbf{Haiku-3.5}} & \multicolumn{2}{c}{\textbf{Llama-3.2-3B}} & \multicolumn{2}{c}{\textbf{Qwen-2.5-Coder}} & \multicolumn{2}{c}{\textbf{DeepSeek-LLM}} & \multicolumn{2}{c}{\textbf{Mistral-7B}} \\

\cmidrule(lr){2-3}\cmidrule(lr){4-5}\cmidrule(lr){6-7}\cmidrule(lr){8-9}\cmidrule(lr){10-11}\cmidrule(lr){12-13}\cmidrule(lr){14-15}\cmidrule(lr){16-17}

 & \makecell{\textit{l}} & \multicolumn{1}{c}{\textit{p}}& \makecell{\textit{l}} & \multicolumn{1}{c}{\textit{p}}& \makecell{\textit{l}} & \multicolumn{1}{c}{\textit{p}}& \makecell{\textit{l}} & \multicolumn{1}{c}{\textit{p}}& \makecell{\textit{l}} & \multicolumn{1}{c}{\textit{p}}& \makecell{\textit{l}} & \multicolumn{1}{c}{\textit{p}}& \makecell{\textit{l}} & \multicolumn{1}{c}{\textit{p}}& \makecell{\textit{l}} & \multicolumn{1}{c}{\textit{p}} \\ \midrule
\makecell{Multi-HumanEval\\\textit{(10 languages*)}} & \makecell[l]{\textbf{Python}\\-} & \makecell[r]{\textbf{100.0\%}\\-} & \makecell[l]{\textbf{Python}\\-} & \makecell[r]{\textbf{100.0\%}\\-} & \makecell[l]{\textbf{Python}\\JavaScript} & \makecell[r]{\textbf{98.8\%}\\1.2\%} & \makecell[l]{\textbf{Python}\\\textit{Total used}} & \makecell[r]{\textbf{99.4\%}\\\textit{4}} & \makecell[l]{\textbf{Python}\\-} & \makecell[r]{\textbf{100.0\%}\\-} & \makecell[l]{\textbf{Python}\\-} & \makecell[r]{\textbf{100.0\%}\\-} & \makecell[l]{\textbf{Python}\\-} & \makecell[r]{\textbf{100.0\%}\\-} & \makecell[l]{\textbf{Python}\\\textit{Total used}} & \makecell[r]{\textbf{100.0\%}\\\textit{3}} \\ \midrule
\makecell{MBXP\\\textit{(10 languages*)}} & \makecell[l]{\textbf{Python}\\-\\-\\-} & \makecell[r]{\textbf{100.0\%}\\-\\-\\-} & \makecell[l]{\textbf{Python}\\JavaScript\\-\\-} & \makecell[r]{\textbf{100.0\%}\\1.0\%\\-\\-} & \makecell[l]{\textbf{Python}\\JavaScript\\Java\\C} & \makecell[r]{\textbf{97.0\%}\\7.5\%\\1.0\%\\0.5\%} & \makecell[l]{\textbf{Python}\\Java\\C++\\JavaScript} & \makecell[r]{\textbf{99.5\%}\\2.5\%\\2.0\%\\2.0\%} & \makecell[l]{\textbf{Python}\\-\\-\\-} & \makecell[r]{\textbf{100.0\%}\\-\\-\\-} & \makecell[l]{\textbf{Python}\\-\\-\\-} & \makecell[r]{\textbf{100.0\%}\\-\\-\\-} & \makecell[l]{\textbf{Python}\\-\\-\\-} & \makecell[r]{\textbf{98.5\%}\\-\\-\\-} & \makecell[l]{\textbf{Python}\\JavaScript\\-\\-} & \makecell[r]{\textbf{99.0\%}\\0.5\%\\-\\-} \\ \midrule
\makecell{AixBench\\\textit{(Java)}} & \makecell[l]{\textbf{Python}\\Java\\JavaScript\\\textit{Total used}} & \makecell[r]{\textbf{75.2\%}\\27.1\%\\10.9\%\\\textit{7}} & \makecell[l]{\textbf{Python}\\Java\\JavaScript\\\textit{Total used}} & \makecell[r]{\textbf{64.3\%}\\38.8\%\\10.1\%\\\textit{7}} & \makecell[l]{\textbf{Java}\\Python\\JavaScript\\\textit{Total used}} & \makecell[r]{\textbf{53.5\%}\\38.8\%\\18.6\%\\\textit{9}} & \makecell[l]{\textbf{Java}\\Python\\JavaScript\\\textit{Total used}} & \makecell[r]{\textbf{59.7\%}\\55.0\%\\20.2\%\\\textit{12}} & \makecell[l]{\textbf{Python}\\Java\\C\#\\\textit{Total used}} & \makecell[r]{\textbf{75.2\%}\\18.6\%\\3.1\%\\\textit{8}} & \makecell[l]{\textbf{Python}\\Java\\JavaScript\\\textit{Total used}} & \makecell[r]{\textbf{68.2\%}\\26.4\%\\3.1\%\\\textit{5}} & \makecell[l]{\textbf{Python}\\Java\\JavaScript\\C\#} & \makecell[r]{\textbf{78.3\%}\\18.6\%\\2.3\%\\0.8\%} & \makecell[l]{\textbf{Python}\\Java\\JavaScript\\\textit{Total used}} & \makecell[r]{\textbf{62.0\%}\\24.8\%\\14.0\%\\\textit{9}} \\ \midrule
\makecell{CoNaLa\\\textit{(Python)}} & \makecell[l]{\textbf{Python}\\JavaScript\\Java\\\textit{Total used}} & \makecell[r]{\textbf{99.0\%}\\6.0\%\\4.5\%\\\textit{7}} & \makecell[l]{\textbf{Python}\\JavaScript\\Java\\C++} & \makecell[r]{\textbf{98.5\%}\\2.5\%\\1.5\%\\0.5\%} & \makecell[l]{\textbf{Python}\\JavaScript\\Java\\\textit{Total used}} & \makecell[r]{\textbf{98.0\%}\\12.5\%\\3.5\%\\\textit{9}} & \makecell[l]{\textbf{Python}\\JavaScript\\Java\\\textit{Total used}} & \makecell[r]{\textbf{99.0\%}\\15.5\%\\13.0\%\\\textit{11}} & \makecell[l]{\textbf{Python}\\JavaScript\\-\\-} & \makecell[r]{\textbf{99.0\%}\\1.0\%\\-\\-} & \makecell[l]{\textbf{Python}\\JavaScript\\Java\\\textit{Total used}} & \makecell[r]{\textbf{98.0\%}\\3.0\%\\2.0\%\\\textit{5}} & \makecell[l]{\textbf{Python}\\JavaScript\\Java\\C++} & \makecell[r]{\textbf{99.0\%}\\1.5\%\\1.0\%\\0.5\%} & \makecell[l]{\textbf{Python}\\JavaScript\\Java\\\textit{Total used}} & \makecell[r]{\textbf{97.5\%}\\4.0\%\\2.5\%\\\textit{5}} \\ \midrule
\makecell{APPS\\\textit{(Python)}} & \makecell[l]{\textbf{Python}\\JavaScript\\Ruby\\R} & \makecell[r]{\textbf{99.5\%}\\1.0\%\\0.5\%\\0.5\%} & \makecell[l]{\textbf{Python}\\JavaScript\\Ruby\\Java} & \makecell[r]{\textbf{99.5\%}\\1.5\%\\0.5\%\\0.5\%} & \makecell[l]{\textbf{Python}\\C++\\JavaScript\\\textit{Total used}} & \makecell[r]{\textbf{93.5\%}\\10.0\%\\5.5\%\\\textit{5}} & \makecell[l]{\textbf{Python}\\C++\\JavaScript\\\textit{Total used}} & \makecell[r]{\textbf{93.5\%}\\7.5\%\\4.5\%\\\textit{6}} & \makecell[l]{\textbf{Python}\\Ruby\\Java\\C} & \makecell[r]{\textbf{98.0\%}\\0.5\%\\0.5\%\\0.5\%} & \makecell[l]{\textbf{Python}\\C++\\Ruby\\R} & \makecell[r]{\textbf{98.0\%}\\1.5\%\\0.5\%\\0.5\%} & \makecell[l]{\textbf{Python}\\Ruby\\-\\-} & \makecell[r]{\textbf{98.5\%}\\0.5\%\\-\\-} & \makecell[l]{\textbf{Python}\\Ruby\\Java\\R} & \makecell[r]{\textbf{98.5\%}\\0.5\%\\0.5\%\\0.5\%} \\ \midrule
\makecell{CodeContests\\\textit{(C++, Java, Python)}} & \makecell[l]{\textbf{Python}\\-\\-} & \makecell[r]{\textbf{100.0\%}\\-\\-} & \makecell[l]{\textbf{Python}\\C++\\-} & \makecell[r]{\textbf{98.5\%}\\2.0\%\\-} & \makecell[l]{\textbf{Python}\\C++\\Java} & \makecell[r]{\textbf{96.5\%}\\19.5\%\\0.5\%} & \makecell[l]{\textbf{Python}\\C++\\-} & \makecell[r]{\textbf{94.0\%}\\15.0\%\\-} & \makecell[l]{\textbf{Python}\\C++\\JavaScript} & \makecell[r]{\textbf{97.5\%}\\2.0\%\\0.5\%} & \makecell[l]{\textbf{Python}\\C++\\-} & \makecell[r]{\textbf{97.0\%}\\6.0\%\\-} & \makecell[l]{\textbf{Python}\\-\\-} & \makecell[r]{\textbf{96.5\%}\\-\\-} & \makecell[l]{\textbf{Python}\\C++\\JavaScript} & \makecell[r]{\textbf{97.5\%}\\2.0\%\\0.5\%} \\ \midrule
\makecell{All datasets} & \makecell[l]{\textbf{Python}\\Java\\JavaScript\\C\#\\\textit{Total used}} & \makecell[r]{\textbf{96.8\%}\\4.0\%\\2.6\%\\0.5\%\\\textit{11}} & \makecell[l]{\textbf{Python}\\Java\\JavaScript\\C++\\\textit{Total used}} & \makecell[r]{\textbf{95.1\%}\\5.0\%\\2.1\%\\0.6\%\\\textit{8}} & \makecell[l]{\textbf{Python}\\Java\\JavaScript\\C++\\\textit{Total used}} & \makecell[r]{\textbf{89.8\%}\\7.3\%\\7.1\%\\6.2\%\\\textit{13}} & \makecell[l]{\textbf{Python}\\Java\\C++\\JavaScript\\\textit{Total used}} & \makecell[r]{\textbf{92.0\%}\\10.3\%\\7.8\%\\6.6\%\\\textit{14}} & \makecell[l]{\textbf{Python}\\Java\\C++\\JavaScript\\\textit{Total used}} & \makecell[r]{\textbf{96.1\%}\\2.3\%\\0.6\%\\0.5\%\\\textit{9}} & \makecell[l]{\textbf{Python}\\Java\\C++\\JavaScript\\\textit{Total used}} & \makecell[r]{\textbf{94.9\%}\\3.5\%\\1.7\%\\0.9\%\\\textit{8}} & \makecell[l]{\textbf{Python}\\Java\\JavaScript\\C\#\\\textit{Total used}} & \makecell[r]{\textbf{96.1\%}\\2.4\%\\0.6\%\\0.1\%\\\textit{6}} & \makecell[l]{\textbf{Python}\\Java\\JavaScript\\C\#\\\textit{Total used}} & \makecell[r]{\textbf{94.1\%}\\3.5\%\\2.6\%\\0.6\%\\\textit{13}} \\ 
\bottomrule
\end{tabular}

    \egroup

    \end{adjustbox}
        \vspace{-2mm}
        \begin{tablenotes}
            \item \tiny{* Multi-HumanEval and MBXP benchmarks contain solutions in the following languages: \textit{C\#, Go, Java, JavaScript, Kotlin, Perl, PHP, Python, Ruby, Scala, Swift, TypeScript}}.
        \end{tablenotes}
    \end{threeparttable}

\end{table*}

\subsection{Experiment 2: Language Preferences}

This experiment investigates LLMs' programming language preferences during code generation.

\paragraph{Benchmark tasks.}
Table~\ref{tab:language-problems} shows the programming languages used by LLMs when generating code for benchmark tasks.
Across nearly all datasets we observe a strong preference for \python.
Except for AixBench, LLMs used \texttt{Python} for at least 93.5\% of tasks, the highest percentage of responses in a non-\texttt{Python} language was considerably lower at 19.5\%.
In the \textit{coding challenge} benchmarks (CodeContests and APPS) chosen for their multilingual solution ecosystems, the LLMs default disproportionately to \python, suggesting that the preference is not merely an artifact of single-language contamination.
The \textit{real world} datasets (AixBench and CoNaLa) show greater language diversity than the synthetic benchmarks: up to twelve different languages appear in responses versus at most five for other datasets. 
Still, the overall set of languages used remains consistently narrow across LLMs.
\textit{Further analysis of the similarities is given in Appendix~\ref{app:res-similar}.}

\paragraph{Project initialisation tasks.}
Table~\ref{tab:language-projects} shows the languages used when LLMs wrote the initial code for various project descriptions where highly-performant languages -- such as \texttt{C}, \texttt{C++} and \texttt{Rust}~\cite{costanzoPerformanceVsProgramming2021} -- are considered optimal.
Even in these settings, the LLMs in our study show a clear preference for using \python: it is the most used language in 23/40 instances, despite being explicitly suboptimal.
Some LLMs are particularly prone to this behaviour (Haiku-3.5, Llama-3.2-3B and DeepSeek-LLM favoured \python for 4/5 projects).
\texttt{JavaScript} is the next most used, but still only the preferred choice in 8/40 instances.

\paragraph{Python analysis.}
Although the preference to \python may be unsurprising, it is still valuable to quantify and understand, for both LLM users and creators.
Bias in an LLM is usually attributed to the training / pre-training stages, where the knowledge of the LLM begins to mirror the underlying data corpora~\cite{guoBiasLargeLanguage2024}.
Therefore, we might expect the \texttt{Python} bias to arise from an abundance of \texttt{Python} in the training data.
The precise code corpora used to train LLMs typically remain undisclosed~\cite{bommasaniFoundationModelTransparency2023}, but there is strong evidence that they are heavily based on large-scale GitHub archives~\cite{majdinasabTrainedMyConsent2025}.

Although \texttt{Python} has seen a sudden surge in popularity, it has not historically been the most used programming language; only in 2024 did it record the highest number of GitHub contributors~\cite{staffOctoverseAILeads2024a}, and lead the TIOBE Index~\cite{jansenTIOBEIndex2025}.
This would suggest that evenly sampling GitHub would yield a more balanced language distribution than the \texttt{Python} preferences we observe.
Therefore, we posit that it is likely that LLM creators are favouring \texttt{Python} when constructing their training corpora; or encouraging \texttt{Python} in the post-training / fine-tuning stages.

\paragraph{\summary{2}}
All LLMs demonstrate a strong preference for \texttt{Python} as their default programming language.
For benchmark tasks, \texttt{Python} accounts for \textbf{90–97\%} of generated solutions.
Even for project initialisation tasks where \texttt{Python} is arguably suboptimal, it remains the most-used language in \textbf{58\%} of cases.

\begin{table*}[ht]
    \caption{\textbf{\textit{Experiment 2: Language Preferences for Project Initialisation Tasks.}} Programming languages used by LLMs when writing project code in scenarios where \python is suboptimal. The languages (\textit{l}) used by each LLM are given, along with the percentage (\textit{p}) of responses that used the language. The most-used language is in bold.}
    \label{tab:language-projects}
    
    \centering
    \begin{adjustbox}{width=\textwidth}
    \bgroup
    \def\arraystretch{1.1}

\begin{tabular}{clrlrlrlrlrlrlrlr}
\toprule
\multirowcell{2}{\textbf{Language}\\\textbf{Task}} & \multicolumn{2}{c}{\textbf{GPT-4o-mini}} & \multicolumn{2}{c}{\textbf{GPT-3.5-turbo}} & \multicolumn{2}{c}{\textbf{Sonnet-3.5}} & \multicolumn{2}{c}{\textbf{Haiku-3.5}} & \multicolumn{2}{c}{\textbf{Llama-3.2-3B}} & \multicolumn{2}{c}{\textbf{Qwen-2.5-Coder}} & \multicolumn{2}{c}{\textbf{DeepSeek-LLM}} & \multicolumn{2}{c}{\textbf{Mistral-7B}} \\

\cmidrule(lr){2-3}\cmidrule(lr){4-5}\cmidrule(lr){6-7}\cmidrule(lr){8-9}\cmidrule(lr){10-11}\cmidrule(lr){12-13}\cmidrule(lr){14-15}\cmidrule(lr){16-17}

 & \makecell{\textit{l}} & \multicolumn{1}{c}{\textit{p}}& \makecell{\textit{l}} & \multicolumn{1}{c}{\textit{p}}& \makecell{\textit{l}} & \multicolumn{1}{c}{\textit{p}}& \makecell{\textit{l}} & \multicolumn{1}{c}{\textit{p}}& \makecell{\textit{l}} & \multicolumn{1}{c}{\textit{p}}& \makecell{\textit{l}} & \multicolumn{1}{c}{\textit{p}}& \makecell{\textit{l}} & \multicolumn{1}{c}{\textit{p}}& \makecell{\textit{l}} & \multicolumn{1}{c}{\textit{p}} \\ \midrule
\makecell{Concurrenct\\web server} & \makecell[l]{\textbf{JavaScript}\\Python\\Go} & \makecell[r]{\textbf{73\%}\\28\%\\2\%} & \makecell[l]{\textbf{JavaScript}\\Python\\-} & \makecell[r]{\textbf{99\%}\\1\%\\-} & \makecell[l]{\textbf{JavaScript}\\-\\-} & \makecell[r]{\textbf{100\%}\\-\\-} & \makecell[l]{\textbf{Python}\\-\\-} & \makecell[r]{\textbf{100\%}\\-\\-} & \makecell[l]{\textbf{Go}\\-\\-} & \makecell[r]{\textbf{100\%}\\-\\-} & \makecell[l]{\textbf{Python}\\-\\-} & \makecell[r]{\textbf{100\%}\\-\\-} & \makecell[l]{\textbf{JavaScript}\\Python\\-} & \makecell[r]{\textbf{94\%}\\6\%\\-} & \makecell[l]{\textbf{JavaScript}\\-\\-} & \makecell[r]{\textbf{100\%}\\-\\-} \\ \midrule
\makecell{Cross-platform\\graphical user\\interface} & \makecell[l]{\textbf{JavaScript}\\Dart\\Python\\C++} & \makecell[r]{\textbf{72\%}\\38\%\\14\%\\3\%} & \makecell[l]{\textbf{JavaScript}\\C++\\Python\\-} & \makecell[r]{\textbf{64\%}\\29\%\\11\%\\-} & \makecell[l]{\textbf{Python}\\Dart\\-\\-} & \makecell[r]{\textbf{99\%}\\2\%\\-\\-} & \makecell[l]{\textbf{Python}\\JavaScript\\TypeScript\\-} & \makecell[r]{\textbf{81\%}\\19\%\\2\%\\-} & \makecell[l]{\textbf{Python}\\-\\-\\-} & \makecell[r]{\textbf{100\%}\\-\\-\\-} & \makecell[l]{\textbf{JavaScript}\\Dart\\-\\-} & \makecell[r]{\textbf{100\%}\\100\%\\-\\-} & \makecell[l]{\textbf{Python}\\-\\-\\-} & \makecell[r]{\textbf{100\%}\\-\\-\\-} & \makecell[l]{\textbf{Dart}\\JavaScript\\-\\-} & \makecell[r]{\textbf{81\%}\\48\%\\-\\-} \\ \midrule
\makecell{Low-latency\\trading\\platform} & \makecell[l]{\textbf{Python}\\-\\-\\-} & \makecell[r]{\textbf{100\%}\\-\\-\\-} & \makecell[l]{\textbf{Python}\\Java\\JavaScript\\C++} & \makecell[r]{\textbf{96\%}\\2\%\\2\%\\1\%} & \makecell[l]{\textbf{Java}\\C++\\Python\\-} & \makecell[r]{\textbf{63\%}\\30\%\\7\%\\-} & \makecell[l]{\textbf{Python}\\-\\-\\-} & \makecell[r]{\textbf{100\%}\\-\\-\\-} & \makecell[l]{\textbf{Python}\\-\\-\\-} & \makecell[r]{\textbf{100\%}\\-\\-\\-} & \makecell[l]{\textbf{Python}\\-\\-\\-} & \makecell[r]{\textbf{100\%}\\-\\-\\-} & \makecell[l]{\textbf{Python}\\-\\-\\-} & \makecell[r]{\textbf{100\%}\\-\\-\\-} & \makecell[l]{\textbf{Python}\\none\\-\\-} & \makecell[r]{\textbf{53\%}\\47\%\\-\\-} \\ \midrule
\makecell{Parallel task\\processing\\library} & \makecell[l]{\textbf{Python}\\C++\\-} & \makecell[r]{\textbf{99\%}\\1\%\\-} & \makecell[l]{\textbf{Python}\\C++\\Java} & \makecell[r]{\textbf{80\%}\\11\%\\9\%} & \makecell[l]{\textbf{Python}\\C++\\-} & \makecell[r]{\textbf{96\%}\\4\%\\-} & \makecell[l]{\textbf{Python}\\-\\-} & \makecell[r]{\textbf{100\%}\\-\\-} & \makecell[l]{\textbf{Python}\\-\\-} & \makecell[r]{\textbf{100\%}\\-\\-} & \makecell[l]{\textbf{C++}\\-\\-} & \makecell[r]{\textbf{100\%}\\-\\-} & \makecell[l]{\textbf{Python}\\-\\-} & \makecell[r]{\textbf{100\%}\\-\\-} & \makecell[l]{\textbf{Python}\\-\\-} & \makecell[r]{\textbf{100\%}\\-\\-} \\ \midrule
\makecell{System-level\\application} & \makecell[l]{\textbf{Python}\\C} & \makecell[r]{\textbf{81\%}\\23\%} & \makecell[l]{\textbf{C}\\Python} & \makecell[r]{\textbf{63\%}\\37\%} & \makecell[l]{\textbf{C}\\Python} & \makecell[r]{\textbf{50\%}\\50\%} & \makecell[l]{\textbf{Python}\\-} & \makecell[r]{\textbf{100\%}\\-} & \makecell[l]{\textbf{Python}\\-} & \makecell[r]{\textbf{100\%}\\-} & \makecell[l]{\textbf{C}\\-} & \makecell[r]{\textbf{100\%}\\-} & \makecell[l]{\textbf{C++}\\-} & \makecell[r]{\textbf{100\%}\\-} & \makecell[l]{\textbf{C++}\\C} & \makecell[r]{\textbf{51\%}\\49\%} \\ 
\bottomrule
\end{tabular}

    \egroup

    \end{adjustbox}
\end{table*}

\begin{table}[ht]

    \caption{\textbf{\textit{Experiment 3: Recommendation Consistency for Project Initialisation Tasks.}} Kendall's $\tau$ correlation between LLMs' NL recommendations and their actual coding choices.
    Only statistically significant correlations are given ($p$-values  $<$ 0.05) - there were none for language tasks.
    Values near 1.0 / -1.0 indicate strong agreement / disagreement.
    }
    \label{tab:internal-consistency}
    
    \centering
    \begin{adjustbox}{width=\columnwidth}
    \bgroup
    \def\arraystretch{1.5}

\begin{tabular}{ccccccccc}
\toprule
\rot{\textbf{Library Task  }} & \rot{\textbf{GPT-4o-mini  }} & \rot{\textbf{GPT-3.5-turbo  }} & \rot{\textbf{Sonnet-3.5  }} & \rot{\textbf{Haiku-3.5  }} & \rot{\textbf{Llama-3.2-3B  }} & \rot{\textbf{Qwen-2.5-Coder  }} & \rot{\textbf{DeepSeek-LLM  }} & \rot{\textbf{Mistral-7B  }} \\ 
\midrule
\makecell{Database} & - & - & - & - & - & - & - & - \\
\rowcolor{lightgray!20} 
Deep learning & - & 0.57 & - & - & - & - & - & - \\
\makecell{Distributed\\computing} & 0.41 & - & 0.60 & 0.60 & - & - & - & - \\
\rowcolor{lightgray!20} 
Web scraper & 0.49 & - & 0.42 & - & - & 0.55 & 0.60 & 0.73 \\
\makecell{Web server} & - & - & - & - & - & - & - & - \\ 
\bottomrule
\end{tabular}

    \egroup

    \end{adjustbox}
\end{table}


    




\subsection{Experiment 3: Recommendation Consistency}

This experiment assesses whether the libraries or programming languages LLMs recommend for a task are consistent with what they actually use.
Table~\ref{tab:internal-consistency} shows the rank correlation results; recommendations are provided in Appendix~\ref{app:res3}.
Overall, we observe very low consistency between the recommended and used technologies.
For library tasks, there is inconsistent moderate agreement ($\tau \approx$ 0.5-0.7): only \textit{distributed computing} and \textit{web scraper} have significant correlations for multiple LLMs, likely because those domains have clearer, community-standard choices.
For language tasks, there is no significant correlation;
direct comparison shows rare agreement, with the top recommended language matching the most used language in only 7/40 instances.
As expected from the task selection, the LLMs did not highly recommend \python (sometimes not suggesting it at all), but still defaulted to using it in most instances.

This inconsistency suggests that LLMs lack a universal knowledge representation of ``which technology best fits these requirements'', which they can apply to both code and NL answers.
This does not imply a specific internal failure mode, but it does indicate that an LLMs' NL answers are a weak signal for what to expect from its generated code.
\textit{In Appendix~\ref{app:reasoning}, we explore prompt engineering strategies as mitigation for this inconsistency.}

\paragraph{\summary{3}}
There is low consistency between the libraries and programming languages LLMs recommend and those that they actually adopt, with the majority of correlations not statistically significant.
For project initialisation tasks, LLMs contradict their own language recommendations \textbf{83\%} of the time.


\section{Discussion}\label{sec:discussion}

Our experiments indicate systematic tendencies in LLMs’ choices of libraries and programming languages.
Here, we balance the practical upsides and downsides of these preferences.

\paragraph{Benefits.}
LLMs' strong tendency to generate code in \python and to rely on widely adopted libraries offers several advantages.
First, \python’s concise syntax and extensive ecosystem improves readability and lowering cognitive load for both novice and expert users.
Second, the reliance on well-established, actively maintained libraries often leads to more reliable and executable outputs.
This behaviour also promotes code standardisation and compatibility across development environments, which can accelerate prototyping, improve reproducibility, and facilitate educational use.

\paragraph{Risks.}
However, this homogenising tendency also introduces several risks.
It may lead to language and library bias, where LLMs systematically overlook alternative languages or emerging libraries that offer better performance, safety, or domain suitability.
Users, especially non-experts, may over-trust the model’s default choices, reproducing outdated or suboptimal dependencies without critical evaluation.
Such bias may distort technology adoption, creating a feedback loop that amplifies the dominance of a few ecosystems represented in training data.

Moreover, excessive reliance on popular libraries can obscure opportunities for innovation and diversity in software design, challenging fairness and inclusivity in software tooling, and reducing discoverability and slow adoption of newer, potentially better tools.
Developers working in under-represented languages or regions may find LLMs less useful or accurate for their contexts.

\paragraph{Practical Implications.}
As an increasing share of code is generated with the help of LLMs, the broader software ecosystem is likely to reflect the coding patterns that LLMs produce --- including their preferred libraries and languages.
At scale, even small systematic preferences can compound, shaping dependency graphs, influencing educational materials, and steering future tooling and benchmarks.

Understanding and measuring these preferences is therefore important to ensure that LLMs makes good technological choices in practice, rather than systematically reinforcing conservative or suboptimal defaults as LLM-generated code becomes more widespread.

The recent rise of coding agents will also shape these dynamics.
As they generate code in real-world repositories, their choices influence ecosystem trends; early evidence suggests more diverse library usage and disciplined dependency addition, potentially mitigating these biases for established repositories~\cite{twistStudyLibraryUsage2026}.



\paragraph{Research challenges.}

We call for future work on understanding, measuring, and mitigating language and library biases in LLM-based code generation, particularly their potential safety implications.
This includes studies that examine how training data distributions shape model preferences, alongside the development of benchmarks and metrics that capture language diversity, dependency selection, and adaptability beyond correctness alone.

Beyond measurement, there is a need to design diversity-aware and evolution-resilient LLMs that maintain proficiency in established languages like \python while supporting innovation and inclusivity across the broader software landscape.
An important complementary direction is the study of other coding preferences -- such as programming paradigms, architectures, data structures, and typing styles -- to determine whether similar biases arise and whether mitigation strategies generalise.


\section{Conclusion}\label{sec:conclusion}

This paper provides the first empirical study on LLM preferences for libraries and programming languages when generating code.
We observe that LLMs consistently favour well-established libraries over high-momentum alternatives;
and exhibit a strong preference towards \python, even for tasks where it can be seen as suboptimal.
Understanding whether these preferences reflect systematic biases -- and when they are helpful versus harmful -- is essential to ensure that future LLMs enrich, rather than narrow, the software ecosystem.


\section*{Limitations \& Threats to Validity}

Here we discuss potential limitations to our study due to the generalisability to other LLMs, and potential threats to internal and external validity.

\paragraph{Generalisability.}
We deliberately use a wide range of production-grade LLMs (open / closed, varying amounts of parameters, and different providers) so that our results reflect common behaviour rather than the quirks of a single LLM.
We observe similar preferences for all LLMs in our study (discussed in further detail in Appendix~\ref{app:res-similar}), the likely root cause of which is that many LLMs are trained or fine-tuned on large GitHub code corpora and \python-centric benchmarks.
Although future (more advanced) LLMs may not exhibit similar tendencies, we believe this is unlikely unless training and evaluation practices change.

\paragraph{Internal validity.}
The threats to internal validity lie in our automatic data extraction and the robustness of our study design.
To alleviate the first threat, we have both unit tested the code responsible and manually verified the extraction process on 100 random samples from language and library experiments.
We took several steps to reduce the second threat and increase confidence in the stability of our results.
Our prompt templates draw on existing studies that generate code with LLMs~\cite{zhuoBigCodeBenchBenchmarkingCode2024} and examine how users interact with LLMs~\cite{nguyenHowBeginningProgrammers2024}; full prompts are reported in Appendix~\ref{app:exp}.
We apply these prompts across multiple task types and domains rather than relying on a single benchmark, which helps ensure that the observed preferences are not tied to a specific task formulation, and we generate multiple responses per prompt to reduce randomness.
We also perform an ablation study (Appendix~\ref{app:res-ablation}) to test sensitivity to prompt style, observing similar patterns regardless of the exact language used.

\paragraph{External validity.}
The threats to external validity lie in datasets and LLMs used, and our analysis of underused libraries.
We mitigate the threat of dataset contamination by carefully selecting benchmark datasets to minimise it, opting for datasets released after the LLMs' knowledge cut-off, or datasets with a variety of programming languages to reduce the possibility that the observed preference is merely caused by contamination.
Another threat is the nondeterministic nature of LLMs and their opaque updates~\cite{sallouBreakingSilenceThreats2024}, introducing variability; we reduce this threat by repeating the experiments multiple times and specifying the exact LLM version to use.
Finally, our case analysis of underused libraries uses GitHub stars as a simple, comparable signal; this inevitably leaves out richer ecosystem cues (downloads, dependents, activity) that future work could use to present a more nuanced view of underused libraries.


\section*{Acknowledgements}
This work is supported by the UKRI Centre for Doctoral Training in Safe and Trusted AI, the EPSRC iCASE Award (ref EP/Y528572/1), the ITEA grants GreenCode (project number 23016), and GENIUS (project number 23026).

\bibliography{main}

\appendix
\appendixpage

\section{LLM Details}
\label{app:llm}

In this study, we evaluate a range of LLMs, with our selection described in Section~\ref{sec:llm}.
To support transparency and reproducibility, Table~\ref{tab:llm} lists the exact version of each LLM and the configuration options used. 

\subsection{API Usage}

Although we rely on default API settings for \textit{temperature} and \textit{top\_p}, we also explicitly set these values during the API calls to ensure reproducibility if default values change in the future.
Closed-source LLMs are prompted using the default values of their corresponding APIs; open-source LLMs are prompted using the default values published on each LLM's \textit{Hugging Face}\footnote{\url{https://huggingface.co/models}} repository.
Each LLM used in this study is accessed via an API, under the providers' terms of service, and is used as expected.

\begin{table*}[t]
    \caption{
        \textbf{\textit{LLM Configuration.}}
        Detailed information of the LLMs used in the study.
        Entries marked with ``--'' indicate that the information was not available at the time of the study (September 2025).
    }
    \label{tab:llm}
    
    \centering
    \begin{adjustbox}{width=\textwidth}
    \bgroup
    \def\arraystretch{1.4}

    \begin{tabular}{cccccccccc}
    \toprule

        \multirowcell{2}[0.25em]{\thead{Model}} & 
        \multirowcell{2}[0.25em]{\thead{Version}} & 
        \multirowcell{2}[0.25em]{\thead{Platform}} &
        \multirowcell{2}[0.25em]{\thead{Release}} & 
        \multirowcell{2}[0.4em]{\thead{Knowledge\\cut-off}} & 
        \multirowcell{2}[0.4em]{\thead{Open-\\source?}} & 
        \multirowcell{2}[0.4em]{\thead{Code\\model?}} & 
        \multirowcell{2}[0.25em]{\thead{Size}} & 
        \multicolumn{2}{c}{\thead{Parameters}} \\
        
         & & & & & & & & \textit{temp} & \textit{top\_p} \\
    \midrule
        GPT-4o-mini~\cite{openaiGPT4TechnicalReport2024} & 
        \texttt{gpt-4o-mini-2024-07-18} & 
        \url{https://openai.com/api/} & 
        July '24 & Oct. '23 & \xmark & \xmark & - & \texttt{1.0} & \texttt{1.0} \\
        
        \rowcolor{lightgray!20} 
        GPT-3.5-turbo~\cite{andrewpengGPT35TurboFinetuning} & 
        \texttt{gpt-3.5-turbo-0125} & 
        \url{https://openai.com/api/} & 
        Nov. '22 & Sep. '21 & \xmark & \xmark & - & \texttt{1.0} & \texttt{1.0} \\
        
        Sonnet-3.5~\cite{anthropicClaude3Model2024} & 
        \texttt{claude-3-5-sonnet-20241022} & 
        \url{https://claude.com/platform/api} & 
        Oct. '24 & July '24 & \xmark & \xmark & - & \texttt{1.0} & \texttt{1.0} \\
        
        \rowcolor{lightgray!20} 
        Haiku-3.5~\cite{anthropicClaude3Model2024} & 
        \texttt{claude-3-5-haiku-20241022} & 
        \url{https://claude.com/platform/api} & 
        Oct. '24 & July '24 & \xmark & \xmark & - & \texttt{1.0} & \texttt{1.0} \\
        
        Llama-3.2-3B~\cite{Llama32Model2025} & 
        \texttt{llama-3.2-3b-instruct-turbo} & 
        \url{https://api.together.xyz/} & 
        Sep. '24 & Dec. '23 & \cmark & \xmark & 3B & \texttt{0.6} & \texttt{0.9} \\
        
        \rowcolor{lightgray!20} 
        Mistral-7B~\cite{jiangMistral7B2023} & 
        \texttt{mistral-7b-instruct-v0.3} & 
        \url{https://api.together.xyz/} & 
        May '24 & May '24 & \cmark & \xmark & 7B & 0.7 & 1.0 \\
        
        Qwen-2.5-Coder~\cite{huiQwen25CoderTechnicalReport2024} & 
        \texttt{qwen-2.5-coder-32b-instruct} & 
        \url{https://api.together.xyz/} & 
        Nov. '24 & Mar. '24 & \cmark & \cmark & 32B & \texttt{0.7} & \texttt{0.8} \\
        
        \rowcolor{lightgray!20} 
        DeepSeek-LLM~\cite{deepseek-aiDeepSeekLLMScaling2024} & 
        \texttt{deepseek-llm-67b-chat} & 
        \url{https://api.together.xyz/} & 
        Nov. '23 & May '23 & \cmark & \xmark & 67B & \texttt{0.7} & \texttt{0.95} \\
    \bottomrule
    \end{tabular}

    \egroup
    \end{adjustbox}
\end{table*}

\subsection{Response Data Extraction}

The LLMs used in the study typically format their responses as \texttt{Markdown} allowing for automatic data extraction, any responses where \texttt{Markdown} cannot be detected are saved separately and subjected to manual analysis for inclusion in our results.
In \texttt{Markdown}, code blocks are denoted by a triple backtick followed by the programming language name~\citep{ExtendedSyntaxMarkdown2025}, therefore, we can use \texttt{regex} matching to extract code blocks and programming languages.
For library experiments, if a \python code block has been extracted, further \texttt{regex} matching is used to extract the imported libraries.
There are two correct syntaxes for external imports in \texttt{Python}~\cite{nzomoAbsoluteVsRelative2025}:
``\texttt{import <library>}'';
and ``\texttt{from <library> import <member>}''.

\section{Experimental Details}
\label{app:exp}

\subsection{Experiment 1: Library Preferences}
\label{app:exp1}

\paragraph{Benchmark Tasks.}

We use BigCodeBench~\cite{zhuoBigCodeBenchBenchmarkingCode2024} as our primary benchmark for investigating library preferences, as discussed in Section~\ref{sub:exp-lib-bench}.
BigCodeBench is released under the Apache license 2.0, allowing usage for this study; and we use the BigCodeBench dataset as intended, to prompt LLMs to generate code.

We use a consistent prompt template for all tasks, inspired by the original BigCodeBench prompt with the added directive to use an external library to ensure that we can best measure the LLM preferences.
The prompt template we use is given in Prompt~\ref{prompt:bcb-template}, and a full example prompt (using a BigCodeBench task from the computation domain) is given in Prompt~\ref{prompt:bcb-example}.

\begin{theorem}\label{prompt:bcb-template}
    ``\textbf{<task description>} You should write self-contained python code. Choose, import and use at least one external library.''
\end{theorem}

\begin{theorem}\label{prompt:bcb-example}
    ``Create a dictionary where keys are specified letters and values are lists of random integers. Then calculate the mean of these integers for each key and return a dictionary of these means. You should write self-contained python code. Choose, import and use at least one external library.''
\end{theorem}

\paragraph{Project Initialisation Tasks.}

To investigate library preferences in a more realistic setting, we use a selection of project initialisation tasks inspired by the groups of libraries on Awesome Python~\cite{pythonAwesomePython2025}, as described in Section~\ref{sub:exp-lib-proj}.
The exact task descriptions used are listed below, along with a selection of the possible libraries for each, as per the libraries given on Awesome Python.

\begin{enumerate}[left=0pt]
    \item \textbf{Database}:
    \textit{``Write the initial python code for a database project with an orm layer.''}
    \\Possible libraries: Django, peewee, Pony, pyDAL, SQLAlchemy.
    
    \item \textbf{Deep learning}:
    \textit{``Write the initial python code for a deep learning project implementing a neural network.''}
    \\Possible libraries: Caffe, Keras, PyTorch, scikit-learn, TensorFlow.
    
    \item \textbf{Distributed computing}:
    \textit{``Write the initial python code for a distributed computing project.''}
    \\Possible libraries: Celery, Dask, Luigi, PySpark, Ray.
    
    \item \textbf{Web scraper}:
    \textit{``Write the initial python code for a web scraping and analysis library.''}
    \\Possible libraries: BeautifulSoup, lxml, MechanicalSoup, Scrapy.
    
    \item \textbf{Web server}:
    \textit{``Write the initial python code for a backend API web server.''}
    \\Possible libraries: Django Rest, FastAPI, Flask, Pyramid, Starlette.
    
\end{enumerate}

\subsection{Experiment 2: Language Preferences}
\label{app:exp2}

\paragraph{Benchmark Tasks.}

We use a selection of benchmark datasets to investigate LLM preferences for programming languages, chosen for their language-agnostic NL-only task descriptions and the likelihood of having solutions published in multiple languages, as described in Section~\ref{sub:exp-lang-bench}.
Here, we provide full details of each of the six datasets chosen, including: the number of tasks; number of valid tasks (those that do not contain references to programming languages); programming languages of the provided ground-truth solutions; and the dataset licence, showing that the dataset is available for use by this study.
Additionally, all dataset usage is consistent with their intended use, to prompt LLMs to generate code.

\begin{enumerate}[left=0pt]
    \item \textbf{Basic}:
    Short coding problems that an entry-level programmer could solve.
    
    \begin{itemize}[left=0pt]
    
        \item \textbf{Multi-HumanEval}~\cite{athiwaratkunMultilingualEvaluationCode2023}:
        A multi-language version of the popular HumanEval~\cite{chenEvaluatingLargeLanguage2021} dataset originally published by OpenAI in 2021, to test code generation from docstrings.
        \\
        \textit{Tasks: 160.}
        \\
        \textit{Solution languages: C\#, Go, Java, JavaScript, Kotlin, Perl, PHP, Python, Ruby, Scala, Swift, Typescript.}
        \\
        \textit{Licence: Apache license 2.0.}
        
        \item \textbf{MBXP}~\cite{athiwaratkunMultilingualEvaluationCode2023}:
        A multi-language version of the popular MBPP~\cite{austinProgramSynthesisLarge2021} dataset originally published by Google in 2021, containing simple problems that require short solutions.
        \\\textit{Tasks: 968.}
        \\
        \textit{Solution languages: C\#, Go, Java, JavaScript, Kotlin, Perl, PHP, Python, Ruby, Scala, Swift, Typescript.}
        \\
        \textit{Licence: Apache license 2.0.}
        
    \end{itemize}
    
    \item \textbf{Real-world}:
    Code generation tasks constructed from realistic coding situations that developers have faced.
    Because the problems are realistic, there are likely to be solutions available online in many languages.
    
    \begin{itemize}[left=0pt]
    
        \item \textbf{AixBench}~\cite{haoAixBenchCodeGeneration2022}: A method-level code generation dataset built from comments found in public GitHub \texttt{Java} methods.
        \\\textit{Tasks: 161 total, 129 valid.}
        \\
        \textit{Solution language: Java.}
        \\
        \textit{Licence: MIT.}
        
        \item \textbf{CoNaLa}~\cite{yinLearningMineAligned2018}: A code generation from NL dataset built from coding problems found on Stack Overflow.
        \\\textit{Tasks: 2,879 total, 2,659 valid.}
        \\
        \textit{Solution language: Python.}
        \\
        \textit{Licence: MIT.}
        
    \end{itemize}
    
    \item \textbf{Coding challenge}:
    In-depth problems that typically require more thinking in order to solve.
    Sourced from online coding competition websites, they are language-agnostic by nature.
    Users can solve the problems in any language they choose, so there are solutions published online in many languages~\cite{furiaCausalAnalysisEmpirical2024}.
    
    \begin{itemize}[left=0pt]
    
        \item \textbf{APPS}~\cite{hendrycksMeasuringCodingChallenge2021}: A large collection of problems from various open-access coding websites, covering various difficulty levels.
        \\\textit{Tasks: 10,000 total, 8,623 valid.}
        \\
        \textit{Solution language: Python.}
        \\
        \textit{Licence: MIT.}
        
        \item \textbf{CodeContests}~\cite{doi:10.1126/science.abq1158}: A competitive coding dataset from 2024, initially used to train AlphaCode~\cite{deepmindCompetitiveProgrammingAlphaCode2024}.
        \\\textit{Tasks: 13,610 total, 12,830 valid.}
        \\
        \textit{Solution languages: C++, Java, Python.}
        \\
        \textit{Licence: Creative Commons Attribution 4.0.}
        
    \end{itemize}
\end{enumerate}

We use a consistent prompt template compatible with each chosen dataset, the exact template is given in Prompt~\ref{prompt:lang-bench}.

\begin{theorem}\label{prompt:lang-bench}
    ``Generate a code-based solution, with an explanation, for the following task or described function: \textbf{<task description>}''
\end{theorem}

\begin{table*}[ht]
    \caption{
    \textbf{\textit{Ground Truth Analysis.}} The \textit{eight} most-used libraries when solving BigCodeBench tasks, with the number of tasks that contain those libraries in the ground-truth solution.
    Per model, $\in$ \textit{GT} shows the number of tasks with a response using that library when it is in the ground-truth, $\notin$ \textit{GT} is the number of tasks with a response using that library when it is \textbf{not} in the ground-truth
    }
    \label{tab:ground-truth}
    
    \centering
    \begin{adjustbox}{width=\textwidth}
    \bgroup
    \def\arraystretch{1.4}

\begin{tabular}{ccrrrrrrrrrrrrrrrr}

\toprule

\multirowcell{2}{\textbf{Library}} & \multirowcell{2}{\textbf{Ground Truth}\\\textbf{Task Count} $\downarrow$} & \multicolumn{2}{c}{\textbf{GPT-4o-mini}} & \multicolumn{2}{c}{\textbf{GPT-3.5-turbo}} & \multicolumn{2}{c}{\textbf{Sonnet-3.5}} & \multicolumn{2}{c}{\textbf{Haiku-3.5}} & \multicolumn{2}{c}{\textbf{Llama-3.2-3B}} & \multicolumn{2}{c}{\textbf{Qwen-2.5-Coder}} & \multicolumn{2}{c}{\textbf{DeepSeek-LLM}} & \multicolumn{2}{c}{\textbf{Mistral-7B}} \\
 & & \makecell{\textit{$\in$ GT}} & \makecell{\textit{$\notin$ GT}} & \makecell{\textit{$\in$ GT}} & \makecell{\textit{$\notin$ GT}} & \makecell{\textit{$\in$ GT}} & \makecell{\textit{$\notin$ GT}} & \makecell{\textit{$\in$ GT}} & \makecell{\textit{$\notin$ GT}} & \makecell{\textit{$\in$ GT}} & \makecell{\textit{$\notin$ GT}} & \makecell{\textit{$\in$ GT}} & \makecell{\textit{$\notin$ GT}} & \makecell{\textit{$\in$ GT}} & \makecell{\textit{$\notin$ GT}} & \makecell{\textit{$\in$ GT}} & \makecell{\textit{$\notin$ GT}} \\ 

\midrule 
pandas & 232 & 197 & 57 & 195 & 57 & 198 & 61 & 197 & 58 & 182 & 52 & 188 & 49 & 167 & 43 & 179 & 43 \\
        \rowcolor{lightgray!20} 
NumPy & 220 & 181 & 133 & 181 & 122 & 188 & 152 & 206 & 165 & 182 & 143 & 155 & 83 & 130 & 76 & 149 & 94 \\

Matplotlib & 216 & 217 & 56 & 215 & 46 & 208 & 63 & 216 & 63 & 214 & 64 & 214 & 50 & 208 & 48 & 207 & 44 \\
        \rowcolor{lightgray!20} 
scikit-learn & 98 & 84 & 4 & 84 & 6 & 80 & 6 & 90 & 5 & 81 & 7 & 88 & 4 & 81 & 6 & 75 & 6 \\

SciPy & 62 & 40 & 8 & 33 & 6 & 42 & 7 & 47 & 12 & 42 & 22 & 41 & 10 & 22 & 7 & 26 & 9 \\
        \rowcolor{lightgray!20} 
Seaborn & 43 & 25 & 10 & 24 & 11 & 28 & 19 & 29 & 28 & 16 & 3 & 22 & 3 & 20 & 3 & 21 & 10 \\

Requests & 25 & 27 & 3 & 27 & 3 & 26 & 2 & 27 & 4 & 26 & 3 & 27 & 2 & 27 & 4 & 27 & 5 \\
        \rowcolor{lightgray!20} 
NLTK & 18 & 10 & 7 & 13 & 6 & 11 & 10 & 12 & 9 & 12 & 12 & 11 & 5 & 7 & 6 & 10 & 10 \\ 
\bottomrule 

\end{tabular}

    \egroup
    \end{adjustbox}
\end{table*}

\paragraph{Project Initialisation Tasks.} \label{sub:language-projects}

To investigate library preferences in a more realistic setting, we use a selection of project initialisation tasks designed to challenge the LLMs' default programming language choice of \python, as described in Section~\ref{sub:exp-lang-proj}.
The exact task descriptions used are listed below, along with a description of why \python is a suboptimal choice compared to other options.

\begin{enumerate}[left=0pt]

    \item \textbf{Concurrent web server}: \textit{``Write the initial code for a high-performance web server to handle a large number of concurrent requests.''}
    \\
    High performance web servers require efficient thread management and concurrency control, areas where \texttt{Rust} (with full async capabilities) or \texttt{C++} (with multithreading) outperform \texttt{Python}~\cite{bugdenSafetyPerformanceProminent2022}.
    
    \item \textbf{Cross-platform graphical user interface}: \textit{``Write the initial code for a modern cross-platform application with a graphical user interface.''}
    \\
    Graphical user interfaces benefit from native performance optimisations and low-latency rendering, \texttt{Python} is inefficient with native libraries, making low-level languages like \texttt{C} and \texttt{C++} more suitable~\cite{karczewskiPythonVsTechnology2021}.
    
    \item \textbf{Low-latency trading platform}: \textit{``Write the initial code for a low-latency trading platform that will allow scaling in the future.''}
    \\
    Financial trading systems demand minimal execution delays, and benefit from precise control over hardware and low-level memory management, areas where \texttt{C++} and \texttt{Go} are better than Python~\cite{nihalRaceZeroLatency2024}.
    
    \item \textbf{Parallel task processing library}: \textit{``Write the initial code for a high performance parallel task processing library.''}
    \\
    \texttt{Python} is suboptimal here because it does not have true parallelism due to the global interpreter lock~\cite{grossPEP703Making}, compiled languages with true built-in parallelism like \texttt{C}, \texttt{C++}, and \texttt{Rust} are more suitable.
    
    \item \textbf{System-level application}: \textit{``Write the initial code for a command line application to perform system-level programming.''}
    \\
    System-level programming inherently relies on direct hardware interaction and efficient memory usage, making \texttt{C}, \texttt{C++}, and \texttt{Rust} better choices than Python~\cite{costanzoPerformanceVsProgramming2021}.
    
\end{enumerate}

\subsection{Experiment 3: Recommendation Consistency}
\label{app:exp3}

For this experiment, we examine whether LLMs are consistent between their library recommendations for a task and what they actually use, as described in Section~\ref{sub:exp-consistency}.
For this, we use Kendall's $\tau$ to compare a \textit{recommendation ranking}, provided by the LLMs NL recommendations, against the \textit{usage ranking} provided by the usage results from Experiments 1 and 2.
Here, we give full details of how we derive both rankings and complete the calculation.

\paragraph{Prompt design.}
We use Prompt~\ref{prompt:rank} to generate LLM recommendations for a task, asking for the best ``python libraries'' or ``languages'' depending on which technological preference we are investigating.
The prompt is intentionally conversational to mirror typical developer-LLM interactions~\cite{akhorozConversationalAICoding2025}; and we do not limit the number of recommendations to allow the LLM to provide all suitable options, as it would be interesting to see if the LLM does not recommend a technology at all but then uses it in a response.
The ranking from each response is manually extracted, with the arithmetic mean rank of each item calculated to give a final ranking.

To check sensitivity to prompt wording, we conducted a small ablation study, varying the exact language used to ask for recommendations, across a handful of representative project initialisation tasks and LLMs.
We used language such as: ``list, in order, the best...'' (final prompt), ``list the top...'', ``recommend the best...''.
The prompt variations produced similar ranked outputs, but we found that asking for the recommendations ``in order'' gave a response that could more reliably be converted into a ranking.

\begin{theorem}\label{prompt:rank}
        “List, in order, the best \textbf{(python libraries | languages)} for the following task: \textbf{<task description>}”
\end{theorem}

\paragraph{Worked example.}
Here we present an example for how we calculate the Kendall $\tau$ statistic for our data, for the LLM GPT-4o-mini and the \textit{concurrent web server} project to investigate library preferences.
We first need to determine the \textit{recommendation ranking} using Prompt~\ref{prompt:rank}.
In our example, the LLM responded with the following three rankings:
\begin{enumerate}[left=0pt]
    \item \texttt{rust, go, javascript, cpp, java, csharp, elixir, python, php, ruby}
    \item \texttt{go, rust, javascript, cpp, java, csharp, elixir, python, kotlin}
    \item \texttt{go, rust, javascript, cpp, csharp, java, python, php, elixir, scala}
\end{enumerate}
Taking averages,  we get a final \textit{recommendation ranking} of: \texttt{go, rust, javascript, cpp, java, csharp, elixir, python, php, kotlin, ruby, scala}

When generating responses for the task in Experiment 2, \texttt{JavaScript} was used 73\% of the time, \python 28\% and \texttt{Go} 2\%.
This gives a \textit{usage ranking} of: \texttt{JavaScript, Python, Go}.

A tuple is then created for each item (e.g. \python, \texttt{Rust}, \texttt{Java}, etc...), containing both of its ranks -- in this example \texttt{JavaScript} has a tuple of (1, 3), and \python has (2, 8).
If an item is not included within a ranking, then it is assumed to be ranked joint last with other items that are also not included.
These pairs of item ranks are then processed using the implementation of Kendall's $\tau$ provided by SciPy~\cite{scipyKendalltauSciPyV1162}, giving both the calculated statistic and the p-value as outputs.

\begin{figure*}
    \centering
    \includegraphics[width=1.0\textwidth]{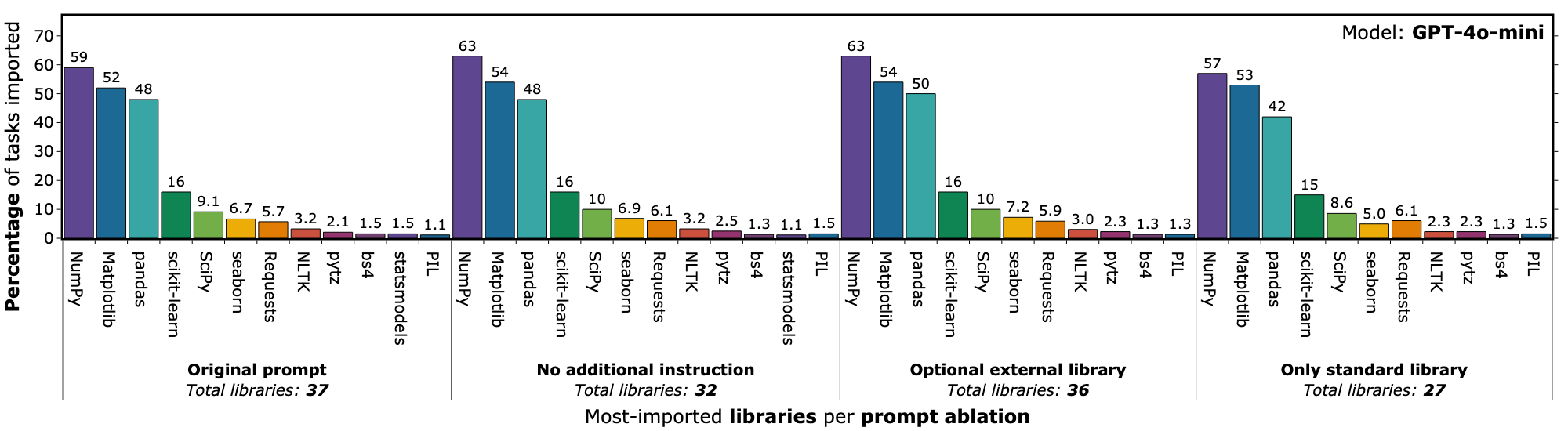}
    \includegraphics[width=1.0\textwidth]{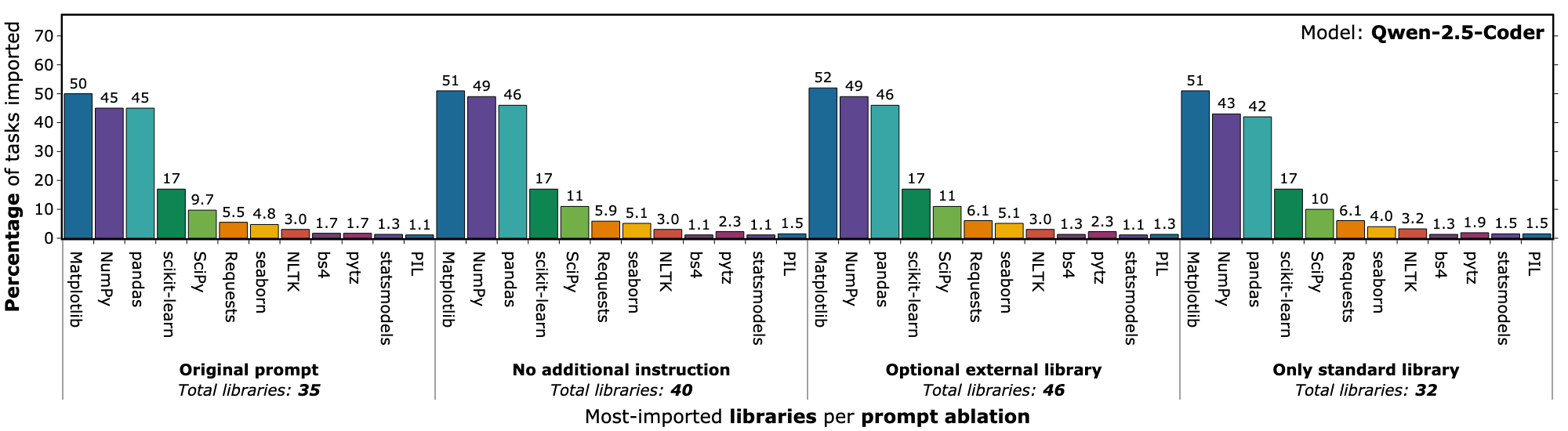}
    \caption{
        \textbf{\textit{Ablation study results.}}
        Libraries used by LLMs when responding to BigCodeBench tasks across various prompt templates.
        For GPT-4o-mini (top) and Qwen-2.5-Coder (bottom), we give the most-used libraries with the percentage of tasks that had a response importing them, and the total unique libraries used.
        Other libraries are imported for \textit{less than 1\%} of tasks.
    }
    \label{fig:ablation}
\end{figure*}

\section{Extended Results}
\label{app:results}

Here we present some additional results that we could not fit into the main paper, along with further statistical analysis of the results from the experiments.


\subsection{Ground Truth Analysis}
\label{app:res1}

In Experiment 1, when investigating LLMs' preferences for libraries when solving benchmark tasks, we observe an apparent overuse of data science libraries; often using a heavyweight library for a simple task (for example, importing \texttt{Numpy} to perform simple calculations).
Here, we investigate this further, presenting some detailed results of how often the most-used libraries are imported in solutions where they match the ground truth.

Table~\ref{tab:ground-truth} shows the results of this analysis.
Although this pattern of overuse is visible in the top three libraries (\texttt{pandas}, \texttt{NumPy}, \texttt{Matplotlib}), we observe that \texttt{NumPy} in particular is used for many tasks where it is not a part of the ground-truth solution.
Across LLMs, for 35-45\% of the tasks for which \texttt{NumPy} is imported, it is not a part of the ground-truth solution.

\begin{table*}[ht]
    \caption{\textbf{\textit{Results for Varying Temperature.}} Languages used for project initialisation tasks when varying temperature. For each temperature (\textit{t}) and project, the preferred languages (\textit{l}) are given, along with the percentage (\textit{p}) of responses that used that language, and a count of the total used (if necessary). The most-used language in each case is in bold.}
    \label{tab:temp-infl}
    
    \centering
    \begin{adjustbox}{width=\textwidth}
    \bgroup
    \def\arraystretch{1.4}

\begin{tabular}{ccccccccccccccccc}
\toprule

\multirowcell{3}{\textbf{Language Task}} & \multicolumn{8}{c}{\textbf{GPT-4o-mini}} & \multicolumn{8}{c}{\textbf{Qwen-2.5-Coder}} \\
\cmidrule(lr){2-9}
\cmidrule(lr){10-17}
& \multicolumn{2}{c}{\textbf{temp=0.0}} & \multicolumn{2}{c}{\textbf{temp=0.5}} & \multicolumn{2}{c}{\textbf{temp=1.0}} & \multicolumn{2}{c}{\textbf{temp=1.5}} & \multicolumn{2}{c}{\textbf{temp=0.0}} & \multicolumn{2}{c}{\textbf{temp=0.5}} & \multicolumn{2}{c}{\textbf{temp=1.0}} & \multicolumn{2}{c}{\textbf{temp=1.5}}\\
& \textit{l} & \multicolumn{1}{c}{\textit{p}} & \textit{l} & \multicolumn{1}{c}{\textit{p}} & \textit{l} & \multicolumn{1}{c}{\textit{p}} & \textit{l} & \multicolumn{1}{c}{\textit{p}} & \textit{l} & \multicolumn{1}{c}{\textit{p}} & \textit{l} & \multicolumn{1}{c}{\textit{p}} & \textit{l} & \multicolumn{1}{c}{\textit{p}} & \textit{l} & \multicolumn{1}{c}{\textit{p}} \\ \midrule

\makecell{Concurrent\\web server} & \makecell[r]{\textbf{JavaScript}\\Python\\Go} & \makecell[l]{\textbf{73\%}\\27\%\\1\%} & \makecell[r]{\textbf{JavaScript}\\Python\\Go} & \makecell[l]{\textbf{63\%}\\36\%\\1\%} & \makecell[r]{\textbf{JavaScript}\\Python\\Go} & \makecell[l]{\textbf{68\%}\\32\%\\3\%} & \makecell[r]{\textbf{JavaScript}\\Python\\\textit{Total used}} & \makecell[l]{\textbf{58\%}\\39\%\\\textit{6}} & \makecell[r]{\textbf{Python}\\Go\\-} & \makecell[l]{\textbf{83\%}\\16\%\\-} & \makecell[r]{\textbf{Python}\\Go\\-} & \makecell[l]{\textbf{85\%}\\15\%\\-} & \makecell[r]{\textbf{Python}\\Go\\JavaScript} & \makecell[l]{\textbf{82\%}\\17\%\\2\%} & \makecell[r]{\textbf{Python}\\Go\\JavaScript} & \makecell[l]{\textbf{86\%}\\11\%\\5\%} \\ \midrule

\makecell{Cross-platform\\graphical user\\interface} & \makecell[r]{\textbf{JavaScript}\\Dart\\\textit{Total used}} & \makecell[l]{\textbf{84\%}\\41\%\\\textit{4}} & \makecell[r]{\textbf{JavaScript}\\Dart\\\textit{Total used}} & \makecell[l]{\textbf{94\%}\\50\%\\\textit{4}} & \makecell[r]{\textbf{JavaScript}\\Dart\\\textit{Total used}} & \makecell[l]{\textbf{76\%}\\48\%\\\textit{5}} & \makecell[r]{\textbf{JavaScript}\\Dart\\\textit{Total used}} & \makecell[l]{\textbf{62\%}\\30\%\\\textit{6}} & \makecell[r]{\textbf{Dart}\\JavaScript\\Python} & \makecell[l]{\textbf{96\%}\\11\%\\3\%} & \makecell[r]{\textbf{Dart}\\JavaScript\\-} & \makecell[l]{\textbf{100\%}\\7\%\\-} & \makecell[r]{\textbf{Dart}\\JavaScript\\Python} & \makecell[l]{\textbf{92\%}\\19\%\\4\%} & \makecell[r]{\textbf{Dart}\\JavaScript\\Python} & \makecell[l]{\textbf{81\%}\\35\%\\2\%} \\ \midrule

\makecell{Low-latency\\trading platform} & \makecell[r]{\textbf{Python}\\C++} & \makecell[l]{\textbf{100\%}\\1\%} & \makecell[r]{\textbf{Python}\\-} & \makecell[l]{\textbf{100\%}\\-} & \makecell[r]{\textbf{Python}\\-} & \makecell[l]{\textbf{100\%}\\-} & \makecell[r]{\textbf{Python}\\JavaScript} & \makecell[l]{\textbf{65\%}\\1\%} & \makecell[r]{\textbf{Python}\\-} & \makecell[l]{\textbf{100\%}\\-} & \makecell[r]{\textbf{Python}\\-} & \makecell[l]{\textbf{100\%}\\-} & \makecell[r]{\textbf{Python}\\-} & \makecell[l]{\textbf{100\%}\\-} & \makecell[r]{\textbf{Python}\\SQL} & \makecell[l]{\textbf{100\%}\\1\%} \\ \midrule

\makecell{Parallel task\\processing library} & \makecell[r]{\textbf{Python}\\C++} & \makecell[l]{\textbf{99\%}\\1\%} & \makecell[r]{\textbf{Python}\\-} & \makecell[l]{\textbf{100\%}\\-} & \makecell[r]{\textbf{Python}\\C++} & \makecell[l]{\textbf{99\%}\\1\%} & \makecell[r]{\textbf{Python}\\C++} & \makecell[l]{\textbf{97\%}\\4\%} & \makecell[r]{\textbf{Python}\\C++} & \makecell[l]{\textbf{58\%}\\42\%} & \makecell[r]{\textbf{Python}\\C++} & \makecell[l]{\textbf{69\%}\\31\%} & \makecell[r]{\textbf{Python}\\C++} & \makecell[l]{\textbf{56\%}\\44\%} & \makecell[r]{\textbf{Python}\\C++} & \makecell[l]{\textbf{62\%}\\38\%} \\ \midrule

\makecell{System-level\\application} & \makecell[r]{\textbf{Python}\\C\\-} & \makecell[l]{\textbf{87\%}\\16\%\\-} & \makecell[r]{\textbf{Python}\\C\\-} & \makecell[l]{\textbf{93\%}\\9\%\\-} & \makecell[r]{\textbf{Python}\\C\\-} & \makecell[l]{\textbf{89\%}\\17\%\\-} & \makecell[r]{\textbf{Python}\\C\\Go} & \makecell[l]{\textbf{83\%}\\17\%\\1\%} & \makecell[r]{\textbf{C}\\-\\-} & \makecell[l]{\textbf{100\%}\\-\\-} & \makecell[r]{\textbf{C}\\-\\-} & \makecell[l]{\textbf{100\%}\\-\\-} & \makecell[r]{\textbf{C}\\-\\-} & \makecell[l]{\textbf{100\%}\\-\\-} & \makecell[r]{\textbf{C}\\-\\-} & \makecell[l]{\textbf{100\%}\\-\\-} \\ \bottomrule

\end{tabular}

    \egroup
    \end{adjustbox}
\end{table*}

\subsection{Ablation Study}
\label{app:res-ablation}

\paragraph{Motivation.}
When analysing library preferences on the benchmark tasks, our main prompt explicitly asks the model to import an external library.
We deliberately made this choice to focus the experiment on library selection, but it raises a fair concern: the instruction itself might inflate external-library usage.
Therefore, to check how sensitive our results are to this wording, we perform a small ablation study.

\paragraph{Experimental Design.}
We repeat the \textit{benchmark task} experiment for library preferences (from Section~\ref{sub:exp-lib-bench}) for both GPT-4o-mini and Qwen-2.5-Coder, using alternative prompt templates.
We keep the BigCodeBench-style request to ``write self-contained python code'', and vary only the additional phrasing so that the requirement to use external libraries is expressed with different levels of strictness.
The prompt templates we use are as follows, from most to least strict regarding the use of extrernal libraries:

\begin{enumerate}[left=0pt]
    \item \textbf{Original prompt:}
    \textit{``\textbf{<task description>} You should write self-contained python code. 
    Choose, import and use at least one external library.''}
    
    \item \textbf{Optional external library:}
    \textit{``\textbf{<task description>} You should write self-contained python code. 
    Use an external library if you want to.''}
    
    \item \textbf{No additional instruction:}
    \textit{``\textbf{<task description>} You should write self-contained python code. 
    ''}
    
    \item \textbf{Only standard library:}
    \textit{``\textbf{<task description>} You should write self-contained python code. 
    Use only the standard library unless strictly necessary.''}
\end{enumerate}

\paragraph{Results.}
The results are shown in Figure~\ref{fig:ablation}.
Across all prompt variants, we observe that the relative library preferences remain remarkably stable.
For both GPT-4o-mini and Qwen-2.5-Coder, the same small set of well-established libraries consistently dominate usage, regardless of whether the prompt explicitly requires an external dependency, merely allows one, or discourages non-standard libraries.
Softening or removing the instruction to use external libraries does not substantially reduce the usage rates of the most common libraries; in some cases, these libraries are used even more frequently.

A clear shift only emerges when the prompt explicitly instructs the LLM to ``use only the standard library.''
Under this condition, both models show a notable reduction in external library usage rates, alongside a decrease in the total number of distinct libraries imported.
But even then, when external libraries are used, they are drawn from the same well-established set observed under other prompt variants.

Overall, these results suggest that prompt phrasing primarily affects how often models rely on external libraries, but has limited influence on which libraries are selected.
The consistency of these patterns across prompt variants and across two different LLMs supports our central finding that LLMs exhibit stable preferences for established libraries, rather than these preferences being an artefact of a particular prompt formulation.

\begin{table}[ht!]
    \tiny
    \caption{\textbf{\textit{LLM Similarities for Benchmark Tasks.}}  Kendall's $\tau$ correlation between different LLMs', comparing the languages and libraries they use for benchmark tasks.
    Only statistically significant correlations are given ($p$-values  $<$ 0.05).
    Values near 1.0 / -1.0 indicate strong agreement / disagreement.}
    \label{tab:model-similarities}
    
    \centering
    \begin{adjustbox}{width=\columnwidth}

    \bgroup
    \def\arraystretch{1.4}

\begin{tabular}{lrrrrrrrrc}
\toprule
 \multicolumn{1}{c}{} & \makecell[c]{\rot{\textbf{GPT-4o-mini  }}} & \makecell[c]{\rot{\textbf{GPT-3.5-turbo  }}} & \makecell[c]{\rot{\textbf{Sonnet-3.5  }}} & \makecell[c]{\rot{\textbf{Haiku-3.5  }}} & \makecell[c]{\rot{\textbf{Llama-3.2-3B  }}} & \makecell[c]{\rot{\textbf{Qwen-2.5-Coder  }}} & \makecell[c]{\rot{\textbf{DeepSeek-LLM  }}} & \makecell[c]{\rot{\textbf{Mistral-7B  }}} & \multicolumn{1}{c}{} \\ 
 \midrule 
\makecell[r]{\textbf{GPT-4o-mini}} & - & 0.66 & 0.58 & 0.49 & 0.44 & - & 0.45 & 0.52 & \multirowcell{8}{\rot{\textbf{Language Tasks}}} \\ 
\makecell[r]{\textbf{GPT-3.5-turbo}} & \cellcolor{lightgray!20}0.58 & - & 0.58 & 0.55 & 0.57 & - & 0.62 & 0.67 &  \\ 
\makecell[r]{\textbf{Sonnet-3.5}} & \cellcolor{lightgray!20}0.49 & \cellcolor{lightgray!20}0.62 & - & 0.50 & 0.60 & - & 0.47 & 0.51 &  \\ 
\makecell[r]{\textbf{Haiku-3.5}} & \cellcolor{lightgray!20}0.50 & \cellcolor{lightgray!20}0.65 & \cellcolor{lightgray!20}0.60 & - & 0.56 & 0.40 & 0.57 & 0.43 &  \\ 
\makecell[r]{\textbf{Llama-3.2-3B}} & \cellcolor{lightgray!20}0.65 & \cellcolor{lightgray!20}0.57 & \cellcolor{lightgray!20}0.50 & \cellcolor{lightgray!20}0.55 & - & 0.51 & 0.53 & 0.65 &  \\ 
\makecell[r]{\textbf{Qwen-2.5-Coder}} & \cellcolor{lightgray!20}0.57 & \cellcolor{lightgray!20}0.62 & \cellcolor{lightgray!20}0.54 & \cellcolor{lightgray!20}0.61 & \cellcolor{lightgray!20}0.51 & - & - & 0.44 &  \\ 
\makecell[r]{\textbf{DeepSeek-LLM}} & \cellcolor{lightgray!20}0.51 & \cellcolor{lightgray!20}0.62 & \cellcolor{lightgray!20}0.44 & \cellcolor{lightgray!20}0.52 & \cellcolor{lightgray!20}0.55 & \cellcolor{lightgray!20}0.47 & - & 0.65 &  \\ 
\makecell[r]{\textbf{Mistral-7B}} & \cellcolor{lightgray!20}0.44 & \cellcolor{lightgray!20}0.55 & \cellcolor{lightgray!20}0.40 & \cellcolor{lightgray!20}0.44 & \cellcolor{lightgray!20}0.49 & \cellcolor{lightgray!20}0.43 & \cellcolor{lightgray!20}0.50 & - &  \\ 
\multicolumn{1}{c}{} & \multicolumn{7}{c}{\cellcolor{lightgray!20}\textbf{Library Tasks}} & \cellcolor{white}{} & \\ 
\bottomrule
\end{tabular}

    \egroup

    \end{adjustbox}
\end{table}

\subsection{Varying Temperature}
\label{app:temperature}

\paragraph{Motivation.}
Temperature is considered the creativity parameter~\cite{peeperkornTemperatureCreativityParameter2024} of LLMs, altering the variability and randomness in responses;
adjusting its value is an obvious method for potentially improving the diversity in LLMs' coding choices.
We also make the deliberate decision in our experiments to use the default temperature for each LLM, to match the expected usage by developers, but this causes the open-source LLMs in our study (that have lower default temperatures) to appear less diverse.
Therefore, we conduct an initial investigation into how adjusting the temperature may allow LLMs to diversify their choice of programming language during project initialisation tasks.

\begin{table*}[ht]
    \caption{\textbf{\textit{LLM Recommendations.}} Top \textit{five} library or programming language recommendations given by each LLM for each project initialisation task.}
    \label{tab:ranks}
    
    \centering
    \begin{adjustbox}{width=\textwidth}
    \bgroup
    \def\arraystretch{1.4}

\begin{tabular}{cllllllll}
\toprule

\thead{Project Task} & \thead{GPT-4o-mini} & \thead{GPT-3.5-turbo} & \thead{Sonnet-3.5} & \thead{Haiku-3.5} & \thead{Llama-3.2-3B} & \thead{Qwen-2.5-Coder} & \thead{DeepSeek-LLM} & \thead{Mistral-7B} \\ \midrule\makecell{Database} & \makecell[l]{1. \texttt{sqlalchemy}\\2. \texttt{djangoorm}\\3. \texttt{peewee}\\4. \texttt{tortoise-orm}\\5. \texttt{ponyorm}\\} & \makecell[l]{1. \texttt{sqlalchemy}\\2. \texttt{peewee}\\3. \texttt{djangoorm}\\4. \texttt{ponyorm}\\5. \texttt{sqlobject}\\} & \makecell[l]{1. \texttt{sqlalchemy}\\2. \texttt{djangoorm}\\3. \texttt{peewee}\\4. \texttt{tortoiseorm}\\5. \texttt{ponyorm}\\} & \makecell[l]{1. \texttt{sqlalchemy}\\2. \texttt{djangoorm}\\3. \texttt{peewee}\\4. \texttt{sqlobject}\\5. \texttt{ponyorm}\\} & \makecell[l]{1. \texttt{sqlalchemy}\\2. \texttt{djangoorm}\\3. \texttt{peewee}\\4. \texttt{ponyorm}\\5. \texttt{djangomodel}\\} & \makecell[l]{1. \texttt{sqlalchemy}\\2. \texttt{djangoorm}\\3. \texttt{peewee}\\4. \texttt{tortoiseorm}\\5. \texttt{sqlobject}\\} & \makecell[l]{1. \texttt{sqlalchemy}\\2. \texttt{djangoorm}\\3. \texttt{peewee}\\4. \texttt{sqlobject}\\5. \texttt{storm}\\} & \makecell[l]{1. \texttt{sqlalchemy}\\2. \texttt{alembic}\\3. \texttt{flask-sqlalchemy}\\4. \texttt{pyramid-sqlalchemy}\\5. \texttt{djangoorm}\\} \\ \midrule
\makecell{Deep\\learning} & \makecell[l]{1. \texttt{tensorflow}\\2. \texttt{torch}\\3. \texttt{keras}\\4. \texttt{fastai}\\5. \texttt{mxnet}\\} & \makecell[l]{1. \texttt{tensorflow}\\2. \texttt{torch}\\3. \texttt{keras}\\4. \texttt{sklearn}\\5. \texttt{numpy}\\} & \makecell[l]{1. \texttt{tensorflow}\\2. \texttt{torch}\\3. \texttt{keras}\\4. \texttt{numpy}\\5. \texttt{sklearn}\\} & \makecell[l]{1. \texttt{tensorflow/keras}\\2. \texttt{torch}\\3. \texttt{jax}\\4. \texttt{mxnet}\\5. \texttt{keras}\\} & \makecell[l]{1. \texttt{numpy}\\2. \texttt{scipy}\\3. \texttt{tensorflow}\\4. \texttt{keras}\\5. \texttt{torch}\\} & \makecell[l]{1. \texttt{tensorflow}\\2. \texttt{torch}\\3. \texttt{keras}\\4. \texttt{fastai}\\5. \texttt{mxnet}\\} & \makecell[l]{1. \texttt{tensorflow}\\2. \texttt{keras}\\3. \texttt{torch}\\4. \texttt{mxnet}\\5. \texttt{chainer}\\} & \makecell[l]{1. \texttt{tensorflow}\\2. \texttt{torch}\\3. \texttt{keras}\\4. \texttt{sklearn}\\5. \texttt{mxnet}\\} \\ \midrule
\makecell{Distributed\\computing} & \makecell[l]{1. \texttt{dask}\\2. \texttt{ray}\\3. \texttt{celery}\\4. \texttt{apachespark}\\5. \texttt{multiprocessing}\\} & \makecell[l]{1. \texttt{dask}\\2. \texttt{celery}\\3. \texttt{pyspark}\\4. \texttt{ray}\\5. \texttt{mpi4py}\\} & \makecell[l]{1. \texttt{ray}\\2. \texttt{dask}\\3. \texttt{pyspark}\\4. \texttt{celery}\\5. \texttt{multiprocessing}\\} & \makecell[l]{1. \texttt{dask}\\2. \texttt{ray}\\3. \texttt{pyspark}\\4. \texttt{celery}\\5. \texttt{zeromq}\\} & \makecell[l]{1. \texttt{dask}\\2. \texttt{joblib}\\3. \texttt{ray}\\4. \texttt{apachespark}\\5. \texttt{mpi4py}\\} & \makecell[l]{1. \texttt{dask}\\2. \texttt{ray}\\3. \texttt{apachespark}\\4. \texttt{celery}\\5. \texttt{joblib}\\} & \makecell[l]{1. \texttt{dask}\\2. \texttt{ray}\\3. \texttt{pyspark}\\4. \texttt{mpi4py}\\5. \texttt{pycompss}\\} & \makecell[l]{1. \texttt{dask}\\2. \texttt{pyspark}\\3. \texttt{celery}\\4. \texttt{ipyparallel}\\5. \texttt{futures}\\} \\ \midrule
\makecell{Web scraper} & \makecell[l]{1. \texttt{requests}\\2. \texttt{bs4}\\3. \texttt{lxml}\\4. \texttt{scrapy}\\5. \texttt{pandas}\\} & \makecell[l]{1. \texttt{bs4}\\2. \texttt{scrapy}\\3. \texttt{pandas}\\4. \texttt{requests}\\5. \texttt{numpy}\\} & \makecell[l]{1. \texttt{requests}\\2. \texttt{bs4}\\3. \texttt{selenium}\\4. \texttt{scrapy}\\5. \texttt{pandas}\\} & \makecell[l]{1. \texttt{requests}\\2. \texttt{bs4}\\3. \texttt{scrapy}\\4. \texttt{lxml}\\5. \texttt{selenium}\\} & \makecell[l]{1. \texttt{bs4}\\2. \texttt{scrapy}\\3. \texttt{requests}\\4. \texttt{lxml}\\5. \texttt{selenium}\\} & \makecell[l]{1. \texttt{requests}\\2. \texttt{bs4}\\3. \texttt{lxml}\\4. \texttt{scrapy}\\5. \texttt{pandas}\\} & \makecell[l]{1. \texttt{bs4}\\2. \texttt{requests}\\3. \texttt{scrapy}\\4. \texttt{selenium}\\5. \texttt{pandas}\\} & \makecell[l]{1. \texttt{requests}\\2. \texttt{bs4}\\3. \texttt{pandas}\\4. \texttt{numpy}\\5. \texttt{matplotlib}\\} \\ \midrule
\makecell{Web server} & \makecell[l]{1. \texttt{flask}\\2. \texttt{fastapi}\\3. \texttt{django}\\4. \texttt{tornado}\\5. \texttt{falcon}\\} & \makecell[l]{1. \texttt{flask}\\2. \texttt{django}\\3. \texttt{fastapi}\\4. \texttt{tornado}\\5. \texttt{sanic}\\} & \makecell[l]{1. \texttt{fastapi}\\2. \texttt{flask}\\3. \texttt{aiohttp}\\4. \texttt{starlette}\\5. \texttt{djangorestframework}\\} & \makecell[l]{1. \texttt{fastapi}\\2. \texttt{flask}\\3. \texttt{djangorestframework}\\4. \texttt{tornado}\\5. \texttt{starlette}\\} & \makecell[l]{1. \texttt{flask}\\2. \texttt{django}\\3. \texttt{fastapi}\\4. \texttt{pyramid}\\5. \texttt{sanic}\\} & \makecell[l]{1. \texttt{flask}\\2. \texttt{djangorestframework}\\3. \texttt{fastapi}\\4. \texttt{tornado}\\5. \texttt{bottle}\\} & \makecell[l]{1. \texttt{flask}\\2. \texttt{django}\\3. \texttt{fastapi}\\4. \texttt{pyramid}\\5. \texttt{bottle}\\} & \makecell[l]{1. \texttt{flask}\\2. \texttt{fastapi}\\3. \texttt{djangorestframework}\\4. \texttt{pyramid}\\5. \texttt{tornado}\\} \\ \midrule
\makecell{Concurrent\\web server} & \makecell[l]{1. \texttt{go}\\2. \texttt{rust}\\3. \texttt{javascript}\\4. \texttt{cpp}\\5. \texttt{java}\\} & \makecell[l]{1. \texttt{go}\\2. \texttt{rust}\\3. \texttt{cpp}\\4. \texttt{java}\\5. \texttt{javascript}\\} & \makecell[l]{1. \texttt{rust}\\2. \texttt{go}\\3. \texttt{cpp}\\4. \texttt{java}\\5. \texttt{javascript}\\} & \makecell[l]{1. \texttt{rust}\\2. \texttt{go}\\3. \texttt{cpp}\\4. \texttt{java}\\5. \texttt{erlang/elixir}\\} & \makecell[l]{1. \texttt{rust}\\2. \texttt{go}\\3. \texttt{cpp}\\4. \texttt{javascript}\\5. \texttt{python}\\} & \makecell[l]{1. \texttt{go}\\2. \texttt{rust}\\3. \texttt{c}\\4. \texttt{cpp}\\5. \texttt{java}\\} & \makecell[l]{1. \texttt{cpp}\\2. \texttt{go}\\3. \texttt{rust}\\4. \texttt{python}\\5. \texttt{javascript}\\} & \makecell[l]{1. \texttt{c/cpp}\\2. \texttt{go}\\3. \texttt{rust}\\4. \texttt{erlang/elixir}\\5. \texttt{java}\\} \\ \midrule
\makecell{Cross-platform\\graphical user\\interface} & \makecell[l]{1. \texttt{javascript}\\2. \texttt{csharp}\\3. \texttt{python}\\4. \texttt{java}\\5. \texttt{dart}\\} & \makecell[l]{1. \texttt{javascript}\\2. \texttt{python}\\3. \texttt{java}\\4. \texttt{dart}\\5. \texttt{csharp}\\} & \makecell[l]{1. \texttt{dart}\\2. \texttt{javascript/typescript}\\3. \texttt{python}\\4. \texttt{csharp}\\5. \texttt{java}\\} & \makecell[l]{1. \texttt{dart}\\2. \texttt{javascript}\\3. \texttt{python}\\4. \texttt{csharp}\\5. \texttt{kotlin}\\} & \makecell[l]{1. \texttt{java}\\2. \texttt{csharp}\\3. \texttt{kotlin}\\4. \texttt{javascript}\\5. \texttt{python}\\} & \makecell[l]{1. \texttt{csharp}\\2. \texttt{kotlin}\\3. \texttt{flutter}\\4. \texttt{react}\\5. \texttt{java}\\} & \makecell[l]{1. \texttt{javascript}\\2. \texttt{python}\\3. \texttt{csharp}\\4. \texttt{java}\\5. \texttt{swift}\\} & \makecell[l]{1. \texttt{javascript}\\2. \texttt{dart}\\3. \texttt{kotlin}\\4. \texttt{swift}\\5. \texttt{csharp}\\} \\ \midrule
\makecell{Low-latency\\trading platform} & \makecell[l]{1. \texttt{cpp}\\2. \texttt{java}\\3. \texttt{rust}\\4. \texttt{go}\\5. \texttt{python}\\} & \makecell[l]{1. \texttt{cpp}\\2. \texttt{java}\\3. \texttt{python}\\4. \texttt{go}\\5. \texttt{rust}\\} & \makecell[l]{1. \texttt{cpp}\\2. \texttt{java}\\3. \texttt{rust}\\4. \texttt{c}\\5. \texttt{go}\\} & \makecell[l]{1. \texttt{cpp}\\2. \texttt{rust}\\3. \texttt{go}\\4. \texttt{java}\\5. \texttt{python}\\} & \makecell[l]{1. \texttt{rust}\\2. \texttt{cpp}\\3. \texttt{go}\\4. \texttt{java}\\5. \texttt{python}\\} & \makecell[l]{1. \texttt{cpp}\\2. \texttt{rust}\\3. \texttt{java}\\4. \texttt{go}\\5. \texttt{python}\\} & \makecell[l]{1. \texttt{cpp}\\2. \texttt{java}\\3. \texttt{python}\\4. \texttt{csharp}\\5. \texttt{javascript}\\} & \makecell[l]{1. \texttt{cpp}\\2. \texttt{java}\\3. \texttt{python}\\4. \texttt{go}\\5. \texttt{rust}\\} \\ \midrule
\makecell{Parallel task\\processing library} & \makecell[l]{1. \texttt{cpp}\\2. \texttt{rust}\\3. \texttt{go}\\4. \texttt{java}\\5. \texttt{python}\\} & \makecell[l]{1. \texttt{cpp}\\2. \texttt{rust}\\3. \texttt{go}\\4. \texttt{java}\\5. \texttt{python}\\} & \makecell[l]{1. \texttt{rust}\\2. \texttt{cpp}\\3. \texttt{go}\\4. \texttt{java}\\5. \texttt{c}\\} & \makecell[l]{1. \texttt{rust}\\2. \texttt{cpp}\\3. \texttt{go}\\4. \texttt{d}\\5. \texttt{julia}\\} & \makecell[l]{1. \texttt{cpp}\\2. \texttt{rust}\\3. \texttt{csharp}\\4. \texttt{java}\\5. \texttt{go}\\} & \makecell[l]{1. \texttt{cpp}\\2. \texttt{rust}\\3. \texttt{go}\\4. \texttt{java}\\5. \texttt{python}\\} & \makecell[l]{1. \texttt{cpp}\\2. \texttt{java}\\3. \texttt{python}\\4. \texttt{csharp}\\5. \texttt{go}\\} & \makecell[l]{1. \texttt{cpp}\\2. \texttt{rust}\\3. \texttt{go}\\4. \texttt{java}\\5. \texttt{scala}\\} \\ \midrule
\makecell{System-level\\application} & \makecell[l]{1. \texttt{c}\\2. \texttt{cpp}\\3. \texttt{rust}\\4. \texttt{go}\\5. \texttt{python}\\} & \makecell[l]{1. \texttt{c}\\2. \texttt{rust}\\3. \texttt{go}\\4. \texttt{cpp}\\5. \texttt{python}\\} & \makecell[l]{1. \texttt{c}\\2. \texttt{rust}\\3. \texttt{cpp}\\4. \texttt{go}\\5. \texttt{assembly}\\} & \makecell[l]{1. \texttt{c}\\2. \texttt{rust}\\3. \texttt{cpp}\\4. \texttt{go}\\5. \texttt{d}\\} & \makecell[l]{1. \texttt{c}\\2. \texttt{cpp}\\3. \texttt{rust}\\4. \texttt{go}\\5. \texttt{assembly}\\} & \makecell[l]{1. \texttt{c}\\2. \texttt{cpp}\\3. \texttt{rust}\\4. \texttt{go}\\5. \texttt{assembly}\\} & \makecell[l]{1. \texttt{c}\\2. \texttt{cpp}\\3. \texttt{rust}\\4. \texttt{go}\\5. \texttt{python}\\} & \makecell[l]{1. \texttt{c/cpp}\\2. \texttt{rust}\\3. \texttt{go}\\4. \texttt{assembly}\\5. \texttt{swift}\\} \\ 

\bottomrule
\end{tabular}

    \egroup
    \end{adjustbox}
\end{table*}

\paragraph{Experimental Design.}
We repeat the \textit{project initialisation} experiment for library preferences (from Section~\ref{sub:exp-lang-proj}) for both an open-source and a closed-source LLM (GPT-4o-mini and Qwen-2.5-Coder), with varying temperatures.
The APIs we use accept temperatures of 0.0-2.0, but larger temperatures can lead to unreliable response parsing~\cite{renzeEffectSamplingTemperature2024}, an effect we observed in our preliminary testing.
Therefore, to ensure reliable response parsing, we use the following temperatures: 0.0, 0.5, 1.0 and 1.5.

\paragraph{Results.}
The results are shown in Table~\ref{tab:temp-infl}.
For GPT-4o-mini, altering temperature has a clear but minimal impact:
for each task, a higher temperature led to the most used language being used in fewer responses, along with a wider variety of languages.
For Qwen-2.5-Coder, altering the temperature seems to have had no effect on the diversity of languages used at all; higher temperatures did not cause a wider variety of languages to be used.
Interestingly, for both LLMs, a temperature of 0.0 does not guarantee that the most used language will remain consistent.

\begin{table*}[ht]
    \tiny
    \caption{\textbf{\textit{Reasoning via Prompt Engineering.}} Languages used for project initialisation when inducing reasoning via prompt engineering. The languages used are given, with the rank assigned to the language by the LLM, and the percentage of responses that used the language for each prompt style. The most-used language in each case has its percentage is in bold. \textit{Base prompt} shows the results from the original experiment, for comparison.}
    \label{tab:cot}
    
    \centering
    \begin{adjustbox}{width=\textwidth}
    \bgroup
    \def\arraystretch{1.4}

\begin{tabular}{ccccccccccccc}
\toprule
\multirowcell{2}{\\\textbf{Language Task}} & \multicolumn{6}{c}{\textbf{GPT-4o}} & \multicolumn{6}{c}{\textbf{Qwen-2.5-Coder}} 
 \\
\cmidrule(lr){2-7}
\cmidrule(lr){8-13}

& \thead{Language} & \thead{Rank $\downarrow$} & \thead{\textit{Base Prompt}} & \thead{Step-by-step} & \thead{Double-check} & \thead{First list} & \thead{Language} & \thead{Rank $\downarrow$} & \thead{\textit{Base Prompt}} & \thead{Step-by-step} & \thead{Double-check} & \thead{First list} \\
\midrule
\makecell{Concurrent\\web server} & \makecell[l]{Go\\Rust\\JavaScript\\Python} & \makecell[r]{\#1\\\#2\\\#3\\\#8} & \makecell[r]{2\%\\0\%\\\textbf{73\%}\\28\%} & \makecell[r]{\textbf{77\%}\\0\%\\23\%\\0\%} & \makecell[r]{47\%\\0\%\\\textbf{53\%}\\0\%} & \makecell[r]{\textbf{98\%}\\2\%\\0\%\\0\%} & \makecell[l]{Go\\Rust\\Python} & \makecell[r]{\#1\\\#2\\\#12} & \makecell[r]{0\%\\0\%\\\textbf{100\%}} & \makecell[r]{\textbf{94\%}\\6\%\\0\%} & \makecell[r]{\textbf{95\%}\\3\%\\2\%} & \makecell[r]{\textbf{92\%}\\8\%\\0\%} \\ \midrule
\makecell{Cross-platform\\graphical user interface} & \makecell[l]{JavaScript\\Python\\Dart\\C++} & \makecell[r]{\#1\\\#3\\\#5\\\#12} & \makecell[r]{\textbf{72\%}\\14\%\\38\%\\3\%} & \makecell[r]{\textbf{84\%}\\9\%\\8\%\\0\%} & \makecell[r]{\textbf{50\%}\\33\%\\21\%\\0\%} & \makecell[r]{\textbf{96\%}\\1\%\\3\%\\0\%} & \makecell[l]{C\#\\Kotlin\\Dart\\Python} & \makecell[r]{\#1\\\#2\\\#6\\\#7} & \makecell[r]{0\%\\0\%\\\textbf{100\%}\\0\%} & \makecell[r]{6\%\\0\%\\\textbf{90\%}\\2\%} & \makecell[r]{4\%\\0\%\\\textbf{81\%}\\15\%} & \makecell[r]{7\%\\2\%\\\textbf{89\%}\\0\%} \\ \midrule
\makecell{Low-latency\\trading platform} & \makecell[l]{C++\\Rust\\Go\\Python} & \makecell[r]{\#1\\\#3\\\#4\\\#5} & \makecell[r]{0\%\\0\%\\0\%\\\textbf{100\%}} & \makecell[r]{\textbf{88\%}\\2\%\\5\%\\9\%} & \makecell[r]{\textbf{66\%}\\12\%\\15\%\\7\%} & \makecell[r]{\textbf{100\%}\\0\%\\0\%\\0\%} & \makecell[l]{C++\\Rust\\Java\\Go\\Python} & \makecell[r]{\#1\\\#2\\\#3\\\#4\\\#5} & \makecell[r]{0\%\\0\%\\0\%\\0\%\\\textbf{100\%}} & \makecell[r]{\textbf{89\%}\\0\%\\1\%\\6\%\\27\%} & \makecell[r]{\textbf{35\%}\\2\%\\0\%\\15\%\\6\%} & \makecell[r]{\textbf{96\%}\\0\%\\0\%\\14\%\\10\%} \\ \midrule
\makecell{Parallel task\\processing library} & \makecell[l]{C++\\Rust\\Go\\Python} & \makecell[r]{\#1\\\#2\\\#3\\\#5} & \makecell[r]{1\%\\0\%\\0\%\\\textbf{99\%}} & \makecell[r]{20\%\\\textbf{54\%}\\20\%\\6\%} & \makecell[r]{8\%\\\textbf{61\%}\\17\%\\15\%} & \makecell[r]{37\%\\\textbf{61\%}\\1\%\\1\%} & \makecell[l]{C++\\Rust\\Python} & \makecell[r]{\#1\\\#2\\\#5} & \makecell[r]{\textbf{100\%}\\0\%\\0\%} & \makecell[r]{\textbf{42\%}\\4\%\\0\%} & \makecell[r]{\textbf{56\%}\\30\%\\4\%} & \makecell[r]{\textbf{66\%}\\34\%\\0\%} \\ \midrule
\makecell{System-level\\application} & \makecell[l]{C\\Rust\\Python} & \makecell[r]{\#1\\\#3\\\#5} & \makecell[r]{23\%\\0\%\\\textbf{81\%}} & \makecell[r]{\textbf{92\%}\\8\%\\4\%} & \makecell[r]{\textbf{96\%}\\5\%\\1\%} & \makecell[r]{\textbf{100\%}\\0\%\\0\%} & \makecell[l]{C\\C++\\Rust} & \makecell[r]{\#1\\\#2\\\#3} & \makecell[r]{\textbf{100\%}\\0\%\\0\%} & \makecell[r]{\textbf{98\%}\\2\%\\0\%} & \makecell[r]{\textbf{94\%}\\4\%\\2\%} & \makecell[r]{\textbf{100\%}\\0\%\\0\%} \\ 

\bottomrule
\end{tabular}

\egroup

\end{adjustbox}
\end{table*}

\subsection{Similarities Across LLMs}
\label{app:res-similar}

From our results LLMs appear to have very similar preferences, here, we do some additional statistical analysis to investigate the extent of this.
We compare the empirical usage rankings of libraries or programming languages for each previous experiment and each pair of LLMs.
We again calculate Kendall's $\tau$ coefficient to understand the rank correlation.
Table~\ref{tab:model-similarities} shows the results for the benchmark tasks.

From the table, we can observe that there is a median coefficient of 0.54 (range 0.40-0.67) across all LLM pairs for languages used for benchmark tasks, with only three results not statistically significant.
For libraries, the coefficients for all LLM pairs are statistically significant, with a median coefficient of 0.53 (range 0.40-0.65).
This indicates that all LLMs have similar preferences when solving benchmark tasks.
For project initialisation tasks, the correlation between preferences is much less clear.
Only 16\% of the coefficients between LLMs have statistical significance, and 13\% is undefined due to an exact match of a single choice of technology.

We assume that the similarity across LLMs reflects shared training data sources -- such as public GitHub repositories~\cite{majdinasabTrainedMyConsent2025} -- while differences arise from their specific data and training variations.
These differences appear to be more pronounced for the open-ended project initialisation tasks, hence the lack of significance in those coefficients.

\subsection{LLM Recommendations}
\label{app:res3}

In Experiment 3 we investigate whether an LLMs NL recommendations for which library or programming language to use for a task aligns with what they actually use when writing code.
In Section~\ref{sec:results} we could only fit the rank correlation results (Table~\ref{tab:internal-consistency}), without details of the \textit{recommendation ranking} itself.
Here, in Table~\ref{tab:ranks}, we present the top five recommendations from all LLMs, for each project initialisation task.
Full rankings are available in our GitHub repository.

\subsection{Reasoning via Prompt Engineering}
\label{app:reasoning}

\paragraph{Motivation.}
Chain-of-thought (CoT) prompting has been shown to elicit reasoning in LLMs~\cite{weiChainofthoughtPromptingElicits2022}; it may enable more optimal programming language choices when coding.
Therefore, we conduct an initial investigation into whether these strategies can improve the consistency between an LLM's programming language recommendations and what it chooses to use when generating code.

\paragraph{Experimental Design.}
We repeat the \textit{project initialisation} experiment for language preferences (from Section~\ref{sub:exp-lang-proj}) for both GPT-4o-mini and Qwen-2.5-Coder.
We append the original prompts with one of the following, either text that has been shown to elicit zero-shot reasoning behaviour, or with the directive to first rank the languages:

\begin{enumerate}[left=0pt]
    \item \textbf{Step-by-step:}
    \textit{``Think step by step about which coding language you should use and why.''}~\cite{kojimaLargeLanguageModels2022}
    \item \textbf{Double-check:}
    \textit{``Double check the reasoning for your coding language choice before writing code.''}~\cite{chowdhuryZeroShotVerificationguidedChain2025}
    \item \textbf{First-list:}
    \textit{``First, list in order, the best coding languages for the task, then use this list to inform your language choice.''}
\end{enumerate}

\paragraph{Results.}
The results are shown in Table~\ref{tab:cot}.
The consistency between NL recommendations and code responses has improved across the board, with the top-ranked programming language now being the most used in 23/30 instances.
Asking the LLM to first rank suitable languages in context shows the best alignment between its original ranking and the languages used, although it was still not perfect, showing potential for variability in the recommendations.
Recommendations for the ``Parallel processing'' task are still inconsistent across all reasoning prompts; both LLMs now use \texttt{Rust} much more, but this only aligned with the recommendations for GPT-4o-mini.
The reasoning prompts seemed to greatly increase the diversity of languages used by Qwen-2.5-Coder, indicating that they create more uncertainty in the responses.

\end{document}